\documentclass[twocolumn,english,american]{revtex4-2}
\usepackage[T1]{fontenc}
\usepackage[utf8]{luainputenc}
\usepackage{amsmath}
\usepackage{amssymb}
\usepackage{graphicx}
\usepackage{babel}
\begin{document}
\title{Energy Window Augmented Plane Waves Approach to Density Functional
Theory}
\author{Garry Goldstein}
\affiliation{garrygoldsteinwinnipeg@gmail.com}
\begin{abstract}
In this work we present a new method for basis set generation for
electronic structure calculations of crystalline solids. This procedure
is aimed at applications to Density Functional Theory (DFT). In this
construction, Energy Window Augmented Plane Waves (EWAPW), we take
advantage of the fact that most DFT calculations use a convergence
loop in order to obtain the self consistent eigenstates of the final
(converged) Kohn Sham (KS) Hamiltonian. Here we propose that, for
the basis used at each step of the self consistency iteration, we
use the previous eigenstate basis, in the interstitial region, and
augment it, inside each Muffin Tin (MT) sphere, with the solution
to the spherically averaged KS Hamiltonian for the linearization energy
of the energy window which contains the energy of that previous eigenstate.
Indeed, to reduce the number of times the spherically averaged KS
potential needs to be solved inside the MT spheres it is advantageous
break up the spectrum into non-overlapping intervals, windows, and
solve the spherically averaged KS Hamiltonian inside the MT region
only once per window per angular momentum channel (at the linearization
energy relevant to that window, usually near the middle of the window).
For practical applications it is reasonable to have on the order of
five to fifty windows. At each step of the iteration of the solution
of the KS equations the EWAPW basis is that of near eigenstates of
the KS Hamiltonian for that iteration. Overall the basis size is the
comparable with the Augmented Plane Waves (APW) basis set but the
number of radial wavefunctions is comparable or greater to Linearized
Augmented Plane Waves + Local Orbitals + Higher Derivative Local Orbitals
+ High Energy Local Orbitals (LAPW+LO+HDLO+HELO) basis set.
\end{abstract}
\maketitle

\section{Introduction}\label{sec:Introduction}

The usefulness of a basis set for all electron calculations of crystalline
solids within Density Functional Theory (DFT) methods is mainly determined
by 1) its linearization error (that is energy errors for eigenstates
of the Kohn Sham (KS) Hamiltonian inside the valence band), 2) its
ability to handle semi-core states - that is the situation when there
is more then one relevant band (not just the valence band) just at
or below the Fermi energy, 3) the effectiveness with which the basis
set converges to the energy obtained in the infinite basis limit as
a function of the number of basis elements for the solution of the
KS equations, $N_{tot}$ - or how hard or soft the basis is, 4) the
efficiently with which one can write down the KS Hamiltonian in that
basis, amongst other things \citep{Singh_2006,Martin_2020,Marx_2009}.
There are now several known competitors, in the literature, for which
basis set is optimal. These can be divided into mostly plane wave
basis methods \citep{Singh_2006,Martin_2020,Singh_1991,Sjostedt_2000,Soler_1989,Soler_1990,Marx_2009,Andersen_1975,Michalicek_2013,Koeling_1975}
and mostly Muffin Tin (MT) basis sets \citep{Martin_2020,Skriver_1984,Andersen_1975,Andersen_1984,Andersen_2003,Korringa_1947,Wills_2010,Kohn_1954,Koeling_1975}.
In this work we focus on the plane wave methods \citep{Singh_2006,Martin_2020,Singh_1991,Sjostedt_2000,Soler_1989,Soler_1990,Marx_2009,Andersen_1975,Michalicek_2013}.
Nominally (without augmentation or pseudization), within the plane
wave method, the basis set is very simple but is extremely inefficient
at obtaining good eigenstates of the KS Hamiltonian for reasonably
small wave vector cutoffs - is extremely hard. Indeed near the nuclei,
where the electron density is high and highly concentrated (oscillating),
the KS potential has high wave vector components so that the eigenstates
of the KS Hamiltonian have high wave vector components as well. This
means that impractically large numbers of plane waves are needed to
accurately describe the environment near atomic nuclei, making the
situation generically numerically impossible. There are two main approaches
to cure this problem - that is still use mainly plane wave basis sets:
pseudopotentials \citep{Vanderbilt_1990,Blochl_1994,Martin_2020,Marx_2009}
and augmentation methods \citep{Andersen_1975,Martin_2020,Michalicek_2014,Singh_1991,Sjostedt_2000,Slater_1937,Smrcka_1984,petru_1985,Michalicek_2013,Koeling_1975}.
In this work we focus on augmentation methods. 

In augmentation methods the main idea is to divide the crystalline
material into two regions: the MT region and the interstitial (IR)
region. The MT region is a set of non-overlapping spheres centered
at the nuclear co-ordinates $\mathbf{r}_{\alpha}$ with radii $S_{\alpha}$
(where $\alpha$ labels the nuclei). The interstitial region is the
rest of the crystal. In the interstitial region a good basis for the
solutions of the KS problem are plane waves - as the KS potential
is varying slowly with position. In the MT region a good basis for
the KS problem is the solutions to the radially averaged KS potential
near each nucleus:
\begin{align}
 & \left[-\frac{d^{2}}{dr^{2}}+\frac{l\left(l+1\right)}{r^{2}}+\overline{V}_{KS}^{n\alpha}\left(r\right)\right]ru_{l\alpha}^{n}\left(r,E\right)=\nonumber \\
 & =Eru_{l\alpha}^{n}\left(r,E\right)\label{eq:Schrodinger}
\end{align}
\foreignlanguage{english}{and $\bar{V}_{KS}^{n\alpha}\left(r\right)$
is the spherically average (KS) potential inside the $\alpha$'th
MT sphere, $n$ is the iteration of the solution to the KS equations
(added into the conventional notation for later convenience), $r=\left|\mathbf{r}-\mathbf{r}_{\alpha}\right|$
is the distance from the center of the MT sphere and $E$ is the energy.}
We note that the electron mass has been set to 1/2. The key question
is what energy, $E$, to use in the solutions of Eq. (\ref{eq:Schrodinger})
when glueing it to the plane waves in the interstitial. 

An early solution to the glueing problem due to Slater was simply
to chose the linearization in the middle of the valence band and augment
with plane waves to obtain the following APW basis $\phi_{\mathbf{k}\mathbf{G}}^{n}\left(E^{n}\right)=$:
\begin{equation}
\left\{ \begin{array}{cc}
\frac{1}{\sqrt{\Omega}}\exp\left(i\left(\mathbf{k}+\mathbf{G}\right)\cdot\mathbf{r}\right) & \mathbf{r}\in IR\\
\sum_{lm}A_{\mathbf{k}\mathbf{G}}^{lm\alpha n}u_{l\alpha}^{n}\left(r,E^{n}\right)Y_{lm}\left(\widehat{\mathbf{r}-\mathbf{r}_{\alpha}}\right) & \mathbf{r}\in MT_{\alpha}
\end{array}\right.\label{eq:APW}
\end{equation}

Where $E^{n}$ is the $n$'th iteration linearization energy, $IR$
is the interstitial, $MT_{\alpha}$ is the $\alpha$th MT sphere,
$\Omega$ is the volume of the unit cell, $\mathbf{k}$ is a point
in the first Brillouin, $\mathbf{G}$ is a reciprocal lattice vector,
$Y_{lm}$ are spherical harmonics, and $A_{\mathbf{k}\mathbf{G}}^{lm\alpha n}$
are the matching coefficients that insure the wavefunction is continuous.
Furthermore the $A_{\mathbf{k}\mathbf{G}}^{lm\alpha n}$ can be computed
using the relation \citep{Loucks_1967}: 
\begin{widetext}
\begin{equation}
\frac{1}{\sqrt{\Omega}}\exp\left(i\left(\mathbf{k}+\mathbf{G}\right)\cdot\mathbf{r}\right)=\frac{4\pi}{\sqrt{\Omega}}\exp\left(i\left(\mathbf{k}+\mathbf{G}\right)\cdot\mathbf{r}_{\alpha}\right)\times\sum_{l,m}Y_{lm}^{*}\left(\widehat{\mathbf{k}+\mathbf{G}}\right)Y_{lm}\left(\widehat{\mathbf{r-\mathbf{r_{\alpha}}}}\right)i^{l}j_{l}\left(\left|\mathbf{k}+\mathbf{G}\right|\left|\mathbf{r}-\mathbf{r}_{\alpha}\right|\right)\label{eq:Famous}
\end{equation}
\end{widetext}

which decouples the matching problem into separate angular momentum
channels. Here $j_{l}$ are the spherical Bessel functions. There
is a cutoff $l_{max}\sim8-12$, which comes about because of the negligibly
small values of the spherical Bessel functions, for large $l$, inside
the MT spheres. This is not an efficient solution because of the large
linearization errors for the wavefunction. Alternatively it is possible
to choose $E^{n}$ self consistently that is the same as every eigenvalue
for each eigenwavefunction (however that is very numerically demanding)
\citep{Martin_2020,Singh_2006}.

Another solution to this glueing problem is the Linearized Augmented
Plane Waves (LAPW) basis set \citep{Andersen_1975,Martin_2020,Singh_2006}.
Here plane waves are augmented within the MT sphere by a linear combination
of the solutions to the spherically averaged KS Hamiltonian (see Eq.
(\ref{eq:Schrodinger})) at the linearization energy $E_{L}^{n}$
(typically the middle of the valence band) $u_{l\alpha}^{n}\left(r,E_{L}^{n}\right)$
and its derivative with respect to energy $\dot{u}_{l\alpha}^{n}\left(r,E_{L}^{n}\right)=\frac{\partial}{\partial E}u_{l\alpha}^{n}\left(r,E_{L}^{n}\right)$.
This is done for all angular momentum channels $lm$ (up to $l_{max}^{L}$)
and all atoms $\alpha$. Whereby, by matching the MT wavefunctions
to the planewave wavefunction, that is demanding the total function
be continuous and continuously differentiable - at the MT sphere radius
\citep{Andersen_1975,Martin_2020,Singh_2006,Soler_1989,Soler_1990}
- we obtain a efficient continuously differentiable basis set. The
LAPW basis set may be written as:
\begin{widetext}
\begin{equation}
\phi_{\mathbf{k}\mathbf{G}}^{Ln}\left(E_{L}^{n}\right)=\left\{ \begin{array}{cc}
\frac{1}{\sqrt{\Omega}}\exp\left(i\left(\mathbf{k}+\mathbf{G}\right)\cdot\mathbf{r}\right) & \mathbf{r}\in IR\\
\sum_{lm}\left[a_{\mathbf{k}\mathbf{G}}^{lm\alpha n}u_{l\alpha}^{n}\left(r,E_{L}^{n}\right)+b_{\mathbf{k}\mathbf{G}}^{lm\alpha n}\dot{u}_{l\alpha}^{n}\left(r,E_{L}^{n}\right)\right]Y_{lm}\left(\widehat{\mathbf{r}-\mathbf{r}_{\alpha}}\right) & \mathbf{r}\in MT_{\alpha}
\end{array}\right.\label{eq:LAPW}
\end{equation}
\end{widetext}

Where $a_{\mathbf{k}\mathbf{G}}^{lm\alpha n}$ and $b_{\mathbf{k}\mathbf{G}}^{lm\alpha n}$
are the matching coefficients that insure the wavefunction is continuous
and continuously differentiable. Furthermore $a_{\mathbf{k}\mathbf{G}}^{lm\alpha n}$
and $b_{\mathbf{k}\mathbf{G}}^{lm\alpha n}$ can be computed using
the relation in Eq. (\ref{eq:Famous}). This allows for efficient
description of the valence band which is adjustable to the form of
the KS potential near the nuclei. This idea can be extended to Linearized
Augmented Lattice Adapted Plane Waves ($\left(LA\right)^{2}PW$) which
have the form \citep{Michalicek_2014}: 
\begin{equation}
\phi_{\mathbf{k}j}^{n}\left(E_{L}^{n}\right)=\sum_{\mathbf{G}}o_{\mathbf{k}\mathbf{G}}^{nj}\phi_{\mathbf{k}\mathbf{G}}^{n}\left(E_{L}^{n}\right)\label{eq:Lattice_adapted}
\end{equation}
(here $j$ labels a basis element). Where one choice for $o_{\mathbf{k}\mathbf{G}}^{nj}$
is given Basis of Early Eigenfunctions $BEE-\left(LA\right)^{2}PW$
where one introduces the solutions to the KS equations \citep{Michalicek_2014}:
\begin{equation}
\psi_{\nu}^{n}=\sum z_{\mathbf{k}\mathbf{G}}^{\nu n}\phi_{\mathbf{k}\mathbf{G}}^{n}\left(E_{L}^{n}\right)\label{eq:Solution}
\end{equation}
 and 
\begin{equation}
o_{\mathbf{k}\mathbf{G}}^{nj}=z_{\mathbf{k}+\mathbf{G}}^{\nu1}\label{eq:Initial}
\end{equation}
for $\nu=j$. One then truncates the basis set with number of basis
functions $j\ll N_{\mathbf{G}}$ (where $N_{\mathbf{G}}$ is the number
of reciprocal lattice points), as one has a basis of near eigenvectors
in the IR, and runs the KS loop to convergence. This greatly reduces
the basis set size needed for the iterative calculations after the
first iteration \citep{Michalicek_2014}. These methods can be further
improved as in Quadratically Augmented Plane Waves QAPW \citep{Smrcka_1984,petru_1985}
where $\ddot{u}_{l\mu}^{n}\left(r,E_{L}^{n}\right)=\frac{\partial^{2}}{\partial E^{2}}u_{l\mu}^{n}\left(r,E_{L}^{n}\right)$
is also used inside the MT sphere and one more derivative (with respect
to the radial co-ordinate) is matched to the plane waves in the IR
which reduces linearization errors. However, the more derivatives
are matched the more the radial wavefunction inside the MT sphere
look like Bessel functions (indeed in the limit where an infinite
number of derivatives is matched the radial wavefunction must be exactly
Bessel) the higher cutoff, the harder the basis becomes \citep{Michalicek_2013,Michalicek_2014}.
Furthermore if there are several relevant bands one can use Local
Orbitals (LO) basis wave functions \citep{Singh_1991,Singh_2006,Sjostedt_2000,Kutepov_2021,Michalicek_2014,Michalicek_2013}.
Many LO like basis wavefunctions are of the form:
\begin{widetext}
\begin{equation}
\phi_{LO}^{nlm\alpha}=\left\{ \begin{array}{cc}
0 & \mathbf{r}\in IR,MT_{\beta}\:for\:\beta\neq\alpha\\
\left[\bar{a}^{l\alpha n}u_{l\alpha}^{n}\left(r,E_{L}^{n}\right)+\bar{b}^{l\alpha n}\dot{u}_{l\alpha}^{n}\left(r,E_{L}^{n}\right)+\bar{c}^{l\alpha n}\tilde{u}_{l\alpha}^{n}\left(r\right)\right]Y_{lm}\left(\widehat{\mathbf{r}-\mathbf{r}_{\alpha}}\right) & \mathbf{r}\in MT_{\alpha}
\end{array}\right.\label{eq:LO}
\end{equation}
\end{widetext}

Where the wavefunction is chosen to be continuous and continuously
differentiable using the coefficients $\bar{a}^{l\alpha n}$, $\bar{b}^{l\alpha n}$
and $\bar{c}^{l\alpha n}$. There are many choices for $\tilde{u}_{l\alpha}^{n}$.
For regular LO these are chosen as $\tilde{u}_{l\alpha}^{n}\left(r\right)=u_{l\alpha}^{n}\left(r,E_{SC}^{n}\right)$
where $SC$ stands for semicore states. For Higher Derivative Local
Orbitals \citep{Kutepov_2021,Betzinger_2011,Michalicek_2013,Michalicek_2014,Firedrich_2006}
(HDLO) $\tilde{u}_{l\alpha}^{n}=\ddot{u}_{l\alpha}^{n}\left(r,E_{L}^{n}\right)$,
while for Higher Energy Local Orbitals \citep{Kutepov_2021,Betzinger_2011,Michalicek_2013,Michalicek_2014,Firedrich_2006}
(HELO) the $\tilde{u}_{l\alpha}^{n}$ is an eigenstate of the spherically
averaged KS Hamiltonian (inside $MT_{\alpha}$) with energy $E_{HELO}^{nl}$
such that 
\begin{equation}
D_{l,k}^{n\alpha}\equiv\frac{S_{\alpha}\frac{\partial}{\partial r}\tilde{u}_{l\alpha}^{n}\left(S_{\alpha},E_{HELO}^{nl}\right)}{\tilde{u}_{l\alpha}^{n}\left(S_{\alpha},E_{HELO}^{nl}\right)}=-\left(l+1\right)\label{eq:HELO}
\end{equation}
Where $\frac{\partial}{\partial r}$ is the radial derivative. This
is chosen to make $\phi_{LO}^{nl\alpha}$ orthogonal to each other
and to the core states. Several HELO bands are possible. This method,
LAPW+LO+HDLO+HELO greatly improves the accuracy of the electronic
structure method.

A competitor to the LAPW method is the APW+lo method which uses the
basis in Eq. (\ref{eq:APW}) and augments it with lo wavefunctions
of the form: 
\begin{widetext}
\[
\phi_{lo}^{nlm\alpha}=\left\{ \begin{array}{cc}
0 & \mathbf{r}\in IR,MT_{\beta}\:for\:\beta\neq\alpha\\
\left[a_{lo}^{l\alpha n}u_{l\alpha}^{n}\left(r,E^{n}\right)+b_{lo}^{l\alpha n}\dot{u}_{l\alpha}^{n}\left(r,E^{n}\right)\right]Y_{lm}\left(\widehat{\mathbf{r}-\mathbf{r}_{\alpha}}\right) & \mathbf{r}\in MT_{\alpha}
\end{array}\right.
\]
\end{widetext}

where the coefficients $a_{lo}^{l\alpha n}$ and $b_{lo}^{l\alpha n}$
are chosen to make the wavefunction continuous. These additional wavefunctions
\citep{Sjostedt_2000,Madsen_2001,Sjostdet_1999} which often come
in a very small number of angular momentum channels \citep{Madsen_2001,Sjostedt_2000,Sjostdet_1999}
lead to a linearization of the APW basis set.

In this work we will not be proposing any new type of LO (such as
HDLO or HELO or conventional LO). As a matter of fact Energy Window
Augmented Plane Wave (EWAPW) can be viewed as a competitor to the
various forms of LO and lo, as we argue that LO and lo type wavefunctions
will not significantly improve the accuracy of the basis set to obtain
the eigenstates of the KS Hamiltonian (see Section \ref{subsec:LO-and-lo}).
Indeed, all of our basis, EWAPW, wavefunctions will be of the form
$\frac{1}{\sqrt{\Omega}}\sum_{\mathbf{G}}\bar{o}_{\mathbf{k}\mathbf{G}}^{nj}\exp\left(i\left(\mathbf{k}+\mathbf{G}\right)\cdot\mathbf{r}\right)$
in the IR, for appropriate $\bar{o}_{\mathbf{k}\mathbf{G}}^{nj}$
as in $\left(LA\right)^{2}PW$. However, unlike $\left(LA\right)^{2}PW$
methods, inside the MT region we will also be modifying the linearization
energy of the $j$'th basis element based on the eigenvalue information
from the previous, $n-1$'st, iteration of the solution of the KS
equations. More precisely $\bar{o}_{\mathbf{k}\mathbf{G}}^{nj}$ will
be chosen so that in the IR the basis wavefunctions have the same
plane wave expansion as eigenstates $\nu$ of the previous iterations
with $\nu=j$. In the MT we will be augmenting these wavefunction
at the energies $E_{j}^{n}$ which are nearly equal to the eigenenergies
of the previous iteration $\varepsilon_{\mathbf{k}\nu}^{n-1}$ (with
$j=\nu$) except we will be using a widowing function $\mathcal{E}_{w}^{n}\left(E\right)$
(which is a piecewise constant approximation to the identity function:
$Id\left(E\right)=E$, see Fig. (\ref{fig:Windowing})) and linearizing
(augmenting) at $E_{j}^{n}=\mathcal{E}_{w}^{n}\left(\varepsilon_{\mathbf{k}\nu}^{n-1}\right)\cong\varepsilon_{\mathbf{k}\nu}^{n-1}$.
As such one needs solve the spherically averaged KS Hamiltonian at
a discrete (on the order of five to fifty) number of values $\mathcal{E}_{w}^{n}\left(E\right)$
takes on - per angular momentum channel, making it practical. Furthermore
as we are able to have many linearization energies the method is likely
more accurate then APW+lo or LAPW (comparable to LAPW+LO+HDLO+HELO),
however we do not increase the basis size beyond APW levels. Detailed
derivations are given in Section \ref{sec:Main-Explanations} below.
We now note that the numerical efficiency of a DFT calculation is
often determined by the diagonalization time and the time it takes
to numerically set up the KS Hamiltonian and charge density \citep{Blugel_2006}.
Overall EWAPW has roughly the same diagonalization time as APW (same
size of basis) but takes a comparable amount of time to set up the
charge density and KS Hamiltonian as LAPW+LO+HDLO+HELO as it has a
comparable number of radial basis functions. We note that in many
situations with a large basis set the diagonalization time is the
biggest \citep{Blugel_2006}.

The rest of this paper is organized as follows. In Section \ref{subsec:Wavefunctions}
we write down the EWAPW wavefunctions in several equivalent ways.
This is the main result of this work. In Section \ref{subsec:Programming-Flow}
we show how to incorporate EWAPW wavefunctions into a KS loop. In
Section \ref{sec:Comments} we will discuss various technical details
needed to make the procedure possible on modern computers as well
as some technical comments useful for computational efficiency. In
Section \ref{sec:EWFLAPW} we present technical steps to write down
the KS Hamiltonian, overlap and electron density, this is highly similar
to regular Full Potential Linearized Augmented Plane Waves (FLAPW).
In Section \ref{sec:Conclusions} we conclude. Various extensions
are considered in the Appendices.

\section{Main Explanations}\label{sec:Main-Explanations}

\subsection{Wavefunctions}\label{subsec:Wavefunctions}

\begin{figure}
\begin{centering}
\includegraphics[width=1\columnwidth]{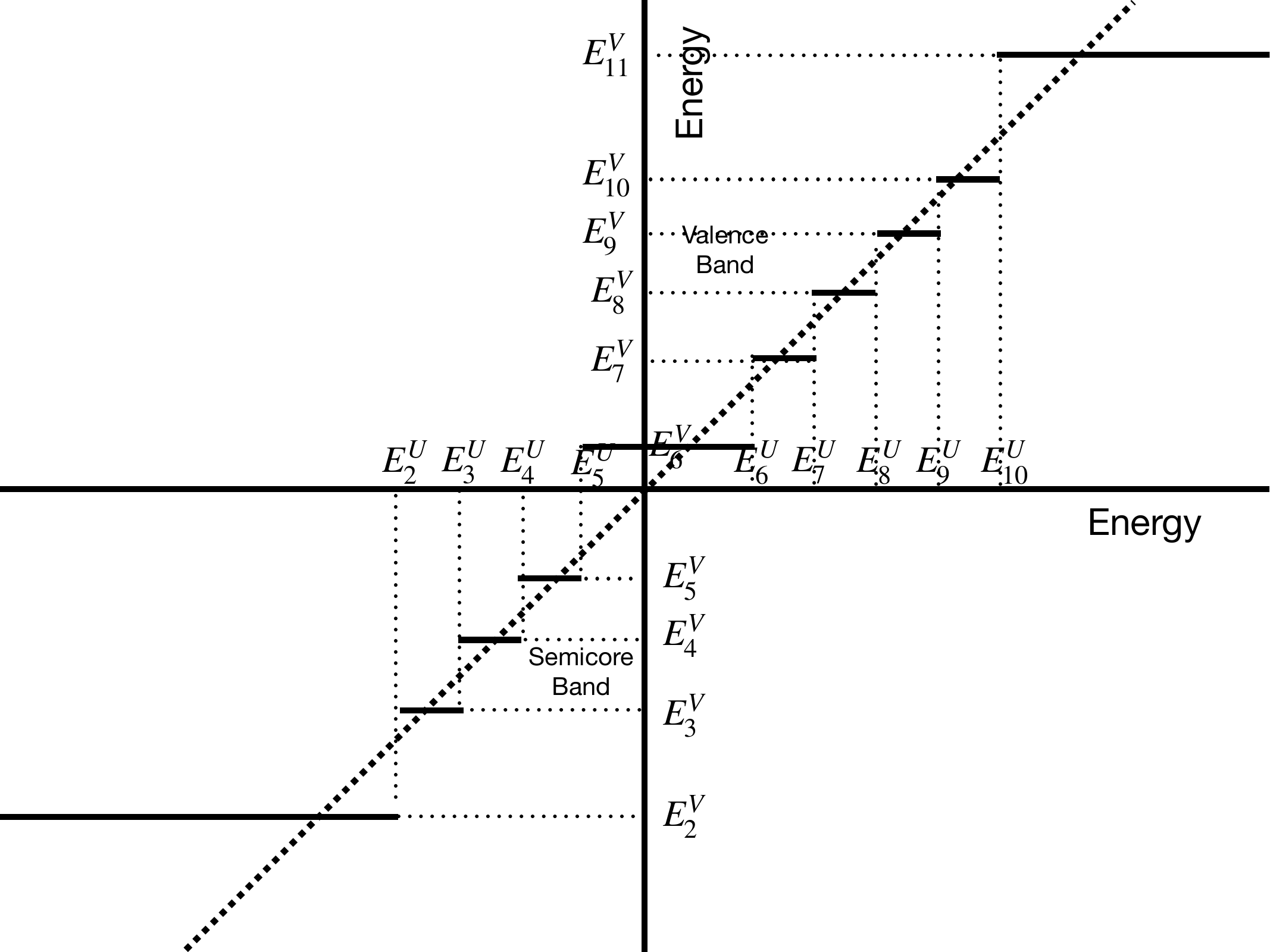}
\par\end{centering}
\caption{The windowing function $\mathcal{E}_{w}\left(E\right)$ (solid lines).
Notice that the steps become more refined near the valence and semicore
bands making $\mathcal{E}_{w}\left(E\right)$ better approximate the
identity function $Id\left(E\right)=E$ in those regions. The identity
function is show for comparison (thick dotted line) and the energies
$E_{i}^{U}$ and $E_{i}^{V}$ are also shown. Here we have suppressed
the index $n$ of the iteration of the solution of the KS equations.}\label{fig:Windowing}
\end{figure}

We now describe in detail the wavefunctions used in the EWAPW basis.
EWAPW is a method to solve the question of which energy $E$ can be
used to glue the wavefunction $u_{l\alpha}^{n}\left(E\right)$ in
the MT to which combination of plane waves in the IR. This is done
using information from the previous iteration of the solutions to
the KS equations. We will not be using and LO/lo (or HDLO or HELO
etc.) wavefunctions and the basis set size, per $\mathbf{k}$ point,
will be the same as the number of reciprocal lattice vectors $N_{\mathbf{G}}$
(or less see Section \ref{subsec:Partial-basis}). We will also avoid
solving for the radially averaged KS equation (see Eq. (\ref{eq:Schrodinger}))
at every energy $\varepsilon_{\mathbf{k}\nu}^{n}$ (which is impractically
hard) by using a windowing function $\mathcal{E}_{w}^{n}\left(E\right)$
that maps nearly via the identity into a small discrete set of values
$\varepsilon_{\mathbf{k}\nu}^{n}\rightarrow E_{i,n}^{V}$ (see Fig.
(\ref{fig:Windowing})) and only solving Eq. (\ref{eq:Schrodinger})
at the energy values $E_{i,n}^{V}$. 

We begin with a definition of a windowing function $\mathcal{E}_{w}^{n}\left(E\right)$,
which is given by: 
\begin{equation}
\mathcal{E}_{w}^{n}\left(E\right)=\sum_{i=2}^{N}E_{i,n}^{V}\cdot\left(\Theta\left(E-E_{i-1,n}^{U}\right)-\Theta\left(E-E_{i,n}^{U}\right)\right)\label{eq:Window}
\end{equation}
Where $E_{1,n}^{U}=-\infty$, $E_{N,n}^{U}=+\infty$ and $E_{i-1,n}^{U}<E_{i,n}^{V}<E_{i,n}^{U}$
$\forall i$. Here $\Theta$ is the Heaviside function. Here $N-1$
, on the order of five to fifty, is the number of windows. The windowing
function is a piecewise constant approximation to the identity function
$Id\left(E\right)=E$. A typical $\mathcal{E}_{w}^{n}\left(E\right)$
is pictured in Fig. (\ref{fig:Windowing}). We note that, as we shall
see below, we will only need to solve Eq. (\ref{eq:Schrodinger})
for the energies $E_{i,n}^{V}$. We further note that the windowing
function divides the spectrum into non-overlapping intervals, windows,
$E_{i-1,n}^{U}\leq E<E_{i,n}^{U}$ and assigns each window the value
$E\rightarrow E_{i,n}^{V}$. Some methods to obtain $E_{i,n}^{V}$
and $E_{i,n}^{U}$ are described in Section (\ref{subsec:How-to-generate}).
Here we simply note that, in general, it is advantageous to place
many $E_{i,n}^{U}$ in the energy range of the valence and semicore
bands (as such approximating the identity more accurately in those
crucial regions) and a few for the rest of the spectrum (which is
less crucial). We now introduce the wavefunctions: $\Phi_{\mathbf{k}\mathbf{G}}^{n}\left(E\right)=$
\begin{equation}
\left\{ \begin{array}{cc}
\frac{1}{\sqrt{\Omega}}\exp\left(i\left(\mathbf{k}+\mathbf{G}\right)\cdot\mathbf{r}\right) & \mathbf{r}\in IR\\
\sum_{lm}\bar{A}_{\mathbf{k}\mathbf{G}}^{lm\alpha n}\left(E\right)u_{l\alpha}^{n}\left(r,\mathcal{E}_{w}^{n}\left(E\right)\right)Y_{lm}\left(\widehat{\mathbf{r}-\mathbf{r}_{\alpha}}\right) & \mathbf{r}\in MT_{\alpha}
\end{array}\right.\label{eq:Windowing_auxhilary}
\end{equation}
Where 
\begin{equation}
\bar{A}_{\mathbf{k}\mathbf{G}}^{lm\alpha n}\left(E\right)=\frac{\frac{1}{\sqrt{\Omega}}4\pi i^{l}j_{l}\left(\left|\mathbf{k}+\mathbf{G}\right|S_{\alpha}\right)Y_{lm}^{*}\left(\widehat{\mathbf{k}+\mathbf{G}}\right)}{u_{l\alpha}^{n}\left(S_{\alpha},\mathcal{E}_{w}^{n}\left(E\right)\right)}\label{eq:Continuous}
\end{equation}
making the wavefunction continuous. We note that: 
\begin{equation}
\Phi_{\mathbf{k}\mathbf{G}}^{n}\left(E\right)=\phi_{\mathbf{k}\mathbf{G}}^{n}\left(\mathcal{E}_{w}^{n}\left(E\right)\right)\label{eq:Def_eqaulity}
\end{equation}
We note that $u_{l\alpha}^{n}\left(r,\mathcal{E}_{w}\left(E\right)\right)=u_{l\alpha}^{n}\left(r,E_{i,n}^{V}\right)$
for some $E_{i,n}^{V}$, so that Eq. (\ref{eq:Schrodinger}) need
only be solved at a discrete (small) number of energy values to obtain
the basis in Eq. (\ref{eq:Windowing_auxhilary}). Now suppose that
for the $n$th iteration of the solutions to the KS equations (see
Eq. (\ref{eq:Secular}) below) we have obtained eigenstates of the
KS equations, $\Psi_{\mathbf{k}\nu}^{n}$ - with eigenvalues $\varepsilon_{\mathbf{k}\nu}^{n}$,
of the form: 
\begin{equation}
\Psi_{\mathbf{k}\nu}^{n}\left(\mathbf{r}\right)=\left\{ \begin{array}{cc}
\sum_{\mathbf{G}}\bar{z}_{\mathbf{k}\mathbf{G}}^{\nu n}\frac{1}{\sqrt{\Omega}}\exp\left(i\left(\mathbf{k}+\mathbf{G}\right)\cdot\mathbf{r}\right) & IR\\
\mathrm{Arbitrary} & MT
\end{array}\right.\label{eq:Basis}
\end{equation}
that is we focus on the expansion of the wavefunction in the IR. Then
we that the EWAPW basis for the $n+1$'st iteration of the solution
of the KS equations is given by:
\begin{equation}
\Psi_{\mathbf{k}j}^{n+1}=\sum_{\mathbf{G}}\bar{z}_{\mathbf{k}\mathbf{G}}^{\nu n}\Phi_{\mathbf{k}\mathbf{G}}^{n+1}\left(\varepsilon_{\mathbf{k}\nu}^{n}\right)\label{eq:Iteration}
\end{equation}
for $j=\nu$ (note the similarities with $\left(LA\right)^{2}PW$
and in particular $BEE-\left(LA\right)^{2}PW$). Or in other words
$\bar{o}_{\mathbf{k}\mathbf{G}}^{n+1j}=\bar{z}_{\mathbf{k}\mathbf{G}}^{\nu n}$
(with $j=\nu$) and we augmenting with $u_{l\alpha}^{n}\left(r,\mathcal{E}_{w}^{n}\left(\varepsilon_{\mathbf{k}\nu}^{n}\right)\right)$
in the MT. In other words we have that the EWAPW basis is given by:
\begin{widetext}
\begin{equation}
\Psi_{\mathbf{k}j}^{n+1}=\left\{ \begin{array}{cc}
\sum_{\mathbf{G}}\bar{z}_{\mathbf{k}\mathbf{G}}^{\nu n}\frac{1}{\sqrt{\Omega}}\exp\left(i\left(\mathbf{k}+\mathbf{G}\right)\cdot\mathbf{r}\right) & \mathbf{r}\in IR\\
\sum_{lm}\left[\sum_{\mathbf{G}}\bar{z}_{\mathbf{k}\mathbf{G}}^{\nu n}A_{\mathbf{k}\mathbf{G}}^{l\alpha n}\left(\mathcal{E}_{w}^{n}\left(\varepsilon_{\mathbf{k}\nu}^{n}\right)\right)\right]\times u_{l\alpha}^{n}\left(r,\mathcal{E}_{w}^{n}\left(\varepsilon_{\mathbf{k}\nu}^{n}\right)\right)\times Y_{lm}\left(\widehat{\mathbf{r}-\mathbf{r}_{\alpha}}\right) & \mathbf{r}\in MT_{\alpha}
\end{array}\right.\label{eq:wave_alternate}
\end{equation}
\end{widetext}

for $j=\nu$. We note that Eqs. (\ref{eq:wave_alternate}) and (\ref{eq:Iteration})
are equivalent but in different notation. Or stated even differently
we have introduced a basis which is given by the previous eigenstate
in the IR and augmented it with the solution of the spherically averaged
KS equations at the energy $\mathcal{E}_{w}^{n+1}\left(\varepsilon_{\mathbf{k}\nu}^{n}\right)$
in the MT regions. 

We notice that the linearization error is at worst given by: 
\begin{equation}
\sim\left(\varepsilon_{\mathbf{k}\nu}^{n+1}-\mathcal{E}_{w}^{n+1}\left(\varepsilon_{\mathbf{k}\nu}^{n}\right)\right)^{2}\label{eq:Linearization}
\end{equation}
When $\mathcal{E}_{w}^{n+1}$ is nearly the identity function $\mathcal{E}_{w}^{n+1}\left(\varepsilon_{\mathbf{k}\nu}^{n}\right)\cong\varepsilon_{\mathbf{k}\nu}^{n}$
and the calculations are well converged $\varepsilon_{\mathbf{k}\nu}^{n+1}\cong\varepsilon_{\mathbf{k}\nu}^{n}$
this error is small, indeed there are multiple $E_{i,n}^{V}$ in the
valence and semicore bands so the error $\left|\varepsilon_{\mathbf{k}\nu}^{n}-\mathcal{E}_{w}^{n+1}\left(\varepsilon_{\mathbf{k}\nu}^{n}\right)\right|$
is much smaller then bandwidth. We note that given that overall we
have created a basis that is given by the previous eigenbasis in the
interstitial and augmented it with the eigenstates of the spherically
averaged KS Hamiltonian (with the linearization energy relevant of
the energies of the previous basis, nearly $\varepsilon_{\mathbf{k}\nu}^{n}$
in the limit of many windows) with good convergence between iterations
and good spherical symmetry in the MT spheres, in EWAPW, we are working
with a basis of near eigenvectors. As such the KS Hamiltonian and
overlap matrices will be nearly diagonal and sparse and as such quicker
to diagonalize then the LAPW Hamiltonian and overlap matrices. We
further note that unlike LAPW or APW+lo we have many radial wavefunctions
$u_{l\alpha}^{n}\left(r,E_{i,n}^{V}\right)$ per angular momentum
channel, it is a comparable or greater amount of radial wavefunctions
to LAPW+LO+HDLO+HELO. Furthermore we note that we only have $N_{\mathbf{G}}$
basis wavefunctions per $\mathbf{k}$ point. Therefore it is reasonable
to surmise that we have a basis with a comparable diagonalization
speed to APW and comparable accuracy to LAPW+LO+HDLO+HELO. We note
that setting up the KS Hamiltonian, overlap and charge density will
take comparable time to setup as in LAPW+LO+HDLO+HELO methods.

\subsection{Programming Flow}\label{subsec:Programming-Flow}

We now show that the code flows, that is the iterative structure of
the KS equation remains intact. Suppose we are passed the following
data from the previous iteration: $\left\{ \rho^{n}\left(\mathbf{r}\right),\varepsilon_{\mathbf{k}\nu}^{n},\bar{z}_{\mathbf{k}\mathbf{G}}^{\nu n}\right\} $
that is the electron density, eigenenergies and the decomposition
of the eigenstates into plane waves in the interstitial. Now we will
show how to pass $\left\{ \rho^{n+1}\left(\mathbf{r}\right),\varepsilon_{\mathbf{k}\nu}^{n+1},\bar{z}_{\mathbf{k}\mathbf{G}}^{\nu n+1}\right\} $
to the iteration after thereby making the code flow. To do so, we
note that if we form the Hamiltonian and overlap matrices $H_{jk}^{\mathbf{k}n}$
and $O_{jk}^{\mathbf{k}n}$ with:
\begin{align}
O_{jk}^{\mathbf{k}n+1} & =\left\langle \Psi_{\mathbf{k}j}^{n+1}\mid\Psi_{\mathbf{k}k}^{n+1}\right\rangle \nonumber \\
H_{jk}^{\mathbf{k}n+1} & =\int d\mathbf{r}\Psi_{\mathbf{k}j}^{n+1}{}^{*}\left(\mathbf{r}\right)\left[-\nabla^{2}+V_{KS}^{n+1}\left(\mathbf{r}\right)\right]\Psi_{\mathbf{k}k}^{n+1}\left(\mathbf{r}\right)\nonumber \\
 & +\mathrm{Boundary\;term}\label{eq:Main_matrices}
\end{align}
 (see Section \ref{sec:EWFLAPW}) then the secular equation is given
by:
\begin{equation}
\sum_{k}H_{jk}^{\mathbf{k}n+1}z_{\mathbf{k}k}^{\nu n+1}=\varepsilon_{\mathbf{k}\nu}^{n+1}\sum_{k}O_{jk}^{\mathbf{k}n+1}\bar{z}_{\mathbf{k}k}^{\nu n+1}\label{eq:Secular}
\end{equation}
then we have that 
\begin{equation}
\bar{z}_{\mathbf{k}\mathbf{G}}^{\nu n+1}=\sum_{k}\bar{z}_{\mathbf{k}k}^{\nu n+1}\bar{z}_{\mathbf{k}\mathbf{G}}^{\mu,n}\label{eq:Matrix}
\end{equation}
Where $\mu=k$ so that we can work in the basis in Eq. (\ref{eq:Iteration})
and obtain the coefficients $\bar{z}_{\mathbf{k}\mathbf{G}}^{\nu n+1}$
needed for the next iteration through matrix multiplication. The electron
density, $\rho^{n+1}\left(\mathbf{r}\right)$, can be obtained from
Section \ref{sec:EWFLAPW}.

\selectlanguage{english}%

\selectlanguage{american}%

\section{Comments}\label{sec:Comments}

There are many minor technical issues that need to be further handled.
Below we describe the solutions to a few of them.

\subsection{Initialization}\label{subsec:Initialization}

The initialization can be handled in a variety of ways. In one method,
the electron density may be chosen as a superposition of electron
densities for the various atoms in the system. The initial linearization
energy may be the same as in the LAPW or APW+lo basis set whereby
we obtain $\left\{ \rho^{1}\left(\mathbf{r}\right),\varepsilon_{\mathbf{k}\nu}^{1},\bar{z}_{\mathbf{k}\mathbf{G}}^{\nu1}\right\} $
and then iterate as in Section \ref{subsec:Programming-Flow}. Alternatively,
we can initialize with the set $\left\{ \rho^{0}\left(\mathbf{r}\right),\varepsilon_{\mathbf{k}\nu}^{0},\bar{z}_{\mathbf{k}\mathbf{G}}^{\nu0}\right\} $
where $\rho^{0}\left(\mathbf{r}\right)$ is the superposition of the
atomic densities and $\nu=\mathbf{G}$ ,with 
\begin{align}
\varepsilon_{\mathbf{k}\nu}^{0} & =\left(\mathbf{k}+\mathbf{G}\right)^{2}+\overline{V}_{KS}^{0}\nonumber \\
\bar{z}_{\mathbf{k}\mathbf{G}'}^{\nu0} & =\delta_{\mathbf{G},\mathbf{G}'}\label{eq:Initialization}
\end{align}
with $\nu=\mathbf{G}$. Here $\overline{V}_{KS}^{0}$ is the spatially
averaged KS potential.

\subsection{Partial Diagonalization }\label{subsec:Partial-Diagonalization}

In many situations it is advantageous for computational speed to only
look for the lowest few bands and not diagonalize the entire KS Hamiltonian
matrix \citep{Singh_2006}. Indeed the higher bands are unoccupied
so do not contribute to the electron density and total energy. In
which case since we are interested in only the lowest bands, say only
$\mathcal{M_{D}}$ of them, we have that the energies $\varepsilon_{\mathbf{k}\nu}^{n},\bar{z}_{\mathbf{k}\mathbf{G}}^{\nu n}$
are computed only with $\nu\leq\mathcal{M_{D}}$. Then for the first
$\mathcal{M_{D}}$ bands we proceed as in Section \ref{subsec:Wavefunctions}.
The rest of the basis may be obtained as in Eq. (\ref{eq:Initialization})
with $0\leftrightarrow n$ and the $\mathbf{G}$'s being chosen such
that the smallest $\mathcal{M}$ energies in Eq. (\ref{eq:Initialization})
are skipped. That is for the higher $N_{\mathbf{G}}-\mathcal{M_{D}}$
bands we write: 
\begin{align}
\varepsilon_{\mathbf{k}\nu}^{n} & =\left(\mathbf{k}+\mathbf{G}\right)^{2}+\overline{V}_{KS}^{n}\nonumber \\
\bar{z}_{\mathbf{k}\mathbf{G}'}^{\nu n} & =\delta_{\mathbf{G},\mathbf{G}'}\label{eq:Initialization-1}
\end{align}
with $\nu=\mathbf{G}$. As such since we have $\left\{ \rho^{n}\left(\mathbf{r}\right),\varepsilon_{\mathbf{k}\nu}^{n},\bar{z}_{\mathbf{k}\mathbf{G}}^{\nu n}\right\} $
we may generate $\left\{ \rho^{n+1}\left(\mathbf{r}\right),\varepsilon_{\mathbf{k}\nu}^{n+1},\bar{z}_{\mathbf{k}\mathbf{G}}^{\nu n+1}\right\} $
and proceed with the iterative solution to the KS equations. This
is not a major loss of accuracy as the KS Hamiltonian and overlap
matrices are nearly diagonal and sparse and the exact for of the high
energy basis does not affect the low energy states significantly. 

\subsection{Expansion of basis set during convergence iteration }\label{subsec:Expansion-of-basis}

In many case it is worthwhile to use small basis sets at the beginning
of the convergence loop and larger basis sets later (see \citep{Goldstein_20024}
and Appendix \ref{sec:Adjusting-the-basis}). In which case we need
to add wavefunctions $\Phi_{\mathbf{k}\mathbf{G}}^{n}\left(E\right)$
with $\mathbf{G}_{min}^{n}<\mathbf{G}<\mathbf{G}_{max}^{n}$. These
can be added to the main procedure in the same way as described in
Section \ref{subsec:Partial-Diagonalization}.

\subsection{Partial basis (compatibility with $BEE-\left(LA\right)^{2}PW$ ideas
\citep{Michalicek_2014})}\label{subsec:Partial-basis}

Here we note that the number of basis elements $\Psi_{\mathbf{k}j}^{n+1}$
may be significantly smaller then the value $N_{\mathbf{G}}$ of reciprocal
wavevectors. Indeed once the system has been initialized it is a basis
of near eigenvectors it is permissible to significantly reduce basis
size. Indeed, as we have a basis of near eigenvectors the Hamiltonian
and overlap matrices are nearly diagonal and sparse. As such it is
permissible to truncate the basis at much smaller basis numbers then
$N_{\mathbf{G}}$. We note that this means that in the interstitial
we limit ourselves to a smaller basis set but one that is adapted
to an earlier IR for an earlier KS Hamiltonian. In principle, to overcome
this last minor difficulty \citep{Michalicek_2014} we can periodically
(every several iterations of the solutions of the KS equations) expand
the basis as in Section \ref{subsec:Expansion-of-basis} to increase
the basis inside the interstitial while still saving on computer time
overall. We will call this method BEE-EWAPW \citep{Michalicek_2014}.

\subsection{EWAPW vs. Energy Window Linearized Augmented Plane Waves (EWLAPW),
multi-radius \citep{Goldstein_20024} options.}\label{subsec:EWAPW-vs.-EWLAPW}

The EWLAPW basis is described in Appendix \ref{sec:EWLAPW} while
multi radius options are described in Appendix \ref{sec:Multi-Radius-options}
(see also ref. \citep{Goldstein_20024}). While these methods are
slightly advantageous over EWAPW - in that they are slightly more
accurate; the added complexity does not seem to be worth the minor
improvement. To see this, we notice that the KS Hamiltonian, $-\nabla^{2}+V_{KS}^{n+1}\left(\mathbf{r}\right)$,
has no singularities on the MT sphere radius. As such the eigenwavefunctions
of the KS Hamiltonian should have no singularities near the surface
of the MT sphere. Since the basis wavefunctions proposed in EWAPW
are near eigenstates they should have no large singularities near
the surface of the MT spheres and should be nearly continuously differentiable
even though the constituent wavefunctions given in Eq. (\ref{eq:Windowing_auxhilary})
have derivative discontinuities. As such adding $\dot{u}_{l\alpha}^{n}\left(r,\mathcal{E}_{w}^{n}\left(\varepsilon_{\mathbf{k}\nu}^{n}\right)\right)$
or varying the radius $S_{\alpha}$ to make the constituent wavefunctions
continuously differentiable would not make much of a difference. From
another perspective it is true that the scaling of the linearization
error improves, we have that the linearization error changes: 
\begin{equation}
\sim\left(\varepsilon_{\mathbf{k}\nu}^{n+1}-\mathcal{E}_{w}^{n+1}\left(\varepsilon_{\mathbf{k}\nu}^{n}\right)\right)^{2}\Rightarrow\sim\left(\varepsilon_{\mathbf{k}\nu}^{n+1}-\mathcal{E}_{w}^{n+1}\left(\varepsilon_{\mathbf{k}\nu}^{n}\right)\right)^{4},\label{eq:Better}
\end{equation}
however it is already very small as the basis is nearly complete see
Eq. (\ref{eq:Linearization}) and the discussion below when the bands
are well split into multiple windows so the improvement of multi radius
and EWLAPW options is minor. 

\subsection{Continuous Interpolation Energy Window Augmented Plane Waves (CEWAPW)}\label{subsec:Continuous-Interpolation-Energy}

We note that CEWAPW is described in Appendix \ref{sec:CEWPAW}. We
note that CEWAPW extensions are not essential in the limit of a large
number of windows but become more important for a moderate number
of windows where they can enhance accuracy at a limited computational
cost. We note that continuous interpolation, multi radius and linearization
options are combinable with each other.

\subsection{LO and lo and related extensions}\label{subsec:LO-and-lo}

We note that LO and lo extensions as well as HDLO and HELO are not
essential. The basis set already has a large number of wavefunctions
at various energies of the spherically averaged KS Hamiltonian inside
the MT spheres so an additional wavefunction with energy close to
those used in the construction of the EWAPW basis would make limited
difference - there are already enough radial wavefunction in the EWAPW
basis. In principle it is possible to remove some windows around the
energies of the LO wavefunctions and use a larger basis involving
LO however that seems disadvantageous. For example, one could remove
all the semicore windows; focus on putting many windows in the valence
band and add some LO, however there are limited advantages to this.
Indeed, this increase of basis set number is not essential for accuracy
but is quite detrimental to computational speed (see Section \ref{sec:Conclusions}).

\subsection{Generating the energy windows}\label{subsec:How-to-generate}

There are many ways to generate energy windows, here we present one
scheme, while many others are possible. The idea of this scheme is
to divide the occupied states into $\mathfrak{N}$ pieces and to choose
the lowest $\mathfrak{M}$ unoccupied bands and divide them into $\mathfrak{P}$
pieces for a total of $\mathfrak{N}+\mathfrak{P}-1$ windows. We begin
by ordering the eigenvalues $\varepsilon_{\mathbf{k}\nu}^{n}\rightarrow\varepsilon^{n}\left(M\right)$
with $\varepsilon^{n}\left(M\right)\leq\varepsilon^{n}\left(M+1\right)$,
$\forall M$. Now we introduce: 
\begin{equation}
N_{O}^{n}=\sum_{\mathbf{k}\nu}\Theta\left(\mu-\varepsilon_{\mathbf{k}\nu}^{n}\right)\label{eq:Occupied}
\end{equation}
which is the number of occupied states. Here $\mu$ is the chemical
potential. Then for $i=2,...\mathfrak{N}$ we have that: 
\begin{align}
E_{i,n+1}^{U} & =\varepsilon^{n}\left(\frac{i}{\mathfrak{N}}N_{O}^{n}\right)\nonumber \\
E_{i,n+1}^{V} & =\frac{\mathfrak{N}}{N_{O}^{n}}\sum_{M=\frac{i-1}{\mathfrak{N}}N_{O}^{n}}^{\frac{i}{\mathfrak{N}}N_{O}^{n}}\varepsilon^{n}\left(M\right)\label{eq:Energy_Upper}
\end{align}
which divide the occupied states into windows (uniformly with respect
to number of states occupied per window) and places the $E_{i,n+1}^{V}$
in the center of mass (with respect to the energy) of those windows.
Here $E_{1,n+1}^{U}=-\infty$. While for $\mathfrak{N}<i<\mathfrak{N}+\mathfrak{P}$
we have that:
\begin{align}
E_{i,n+1}^{U} & =\varepsilon^{n}\left(N_{O}^{n}+\frac{i-\mathfrak{N}}{\mathfrak{P}}\left(\mathfrak{M}\cdot N_{\mathbf{k}}\right)\right)\nonumber \\
E_{i,n+1}^{V} & =\frac{\mathfrak{P}}{\mathfrak{M}\cdot N_{\mathbf{k}}}\sum_{M=N_{O}^{n}+\frac{i-\mathfrak{N}-1}{\mathfrak{P}}\left(\mathfrak{M}\cdot N_{\mathbf{k}}\right)}^{M=N_{O}^{n}+\frac{i-\mathfrak{N}}{\mathfrak{P}}\left(\mathfrak{M}\cdot N_{\mathbf{k}}\right)}\varepsilon^{n}\left(M\right)\label{eq:Energy_value}
\end{align}
which divide the $\mathfrak{M}$ lowest unoccupied states (uniformly
with respect to number of states occupied per window) and places the
$E_{i,n+1}^{V}$ in the center of mass (with respect to the energy)
of those windows. Where
\begin{equation}
N_{\mathbf{k}}=\sum_{\mathbf{k}}1\label{eq:N_k}
\end{equation}
is the number of $\mathbf{k}$ points. With 
\begin{equation}
E_{\mathfrak{N}+\mathfrak{P},n+1}^{V}=\frac{\mathfrak{P}}{\mathfrak{M}\cdot N_{\mathbf{k}}}\sum_{M=N_{O}^{n}+\frac{\mathfrak{P}-1}{\mathfrak{P}}\left(\mathfrak{M}\cdot N_{\mathbf{k}}\right)}^{M=N_{O}^{n}+\left(\mathfrak{M}\cdot N_{\mathbf{k}}\right)}\varepsilon^{n}\left(M\right)\label{eq:Weight}
\end{equation}
while $E_{\mathfrak{N}+\mathfrak{P},n+1}^{U}=+\infty$. Whereby we
obtain a specific scheme for generation $E_{i,n+1}^{U}$ and $E_{i,n+1}^{V}$,
others are also possible.

\subsection{Reducing the number of angular momentum channels}\label{subsec:Reducing-the-number}
\begin{widetext}
Since the EWAPW basis is a basis of near eigenstates of the KS Hamiltonian,
the combinations $\sum_{\mathbf{G}}\bar{z}_{\mathbf{k}\mathbf{G}}^{\nu n}\exp\left(i\left(\mathbf{k}+\mathbf{G}\right)\cdot\mathbf{r}\right)$
are given by:

\begin{equation}
\sum_{\mathbf{G}}\frac{\bar{z}_{\mathbf{k}\mathbf{G}}^{\nu n}}{\sqrt{\Omega}}\exp\left(i\left(\mathbf{k}+\mathbf{G}\right)\cdot\mathbf{r}\right)=4\pi\sum_{l,m}i^{l}\left[\sum_{\mathbf{G}}\frac{\bar{z}_{\mathbf{k}\mathbf{G}}^{\nu n}}{\sqrt{\Omega}}Y_{lm}^{*}\left(\widehat{\mathbf{k}+\mathbf{G}}\right)j_{l}\left(\left|\mathbf{k}+\mathbf{G}\right|\left|\mathbf{r}-\mathbf{r}_{\alpha}\right|\right)\exp\left(i\left(\mathbf{k}+\mathbf{G}\right)\cdot\mathbf{r}_{\alpha}\right)\right]Y_{lm}\left(\widehat{\mathbf{r-\mathbf{r_{\alpha}}}}\right)\label{eq:Famous-1}
\end{equation}
Whereby we have that
\begin{equation}
\sum_{\mathbf{G}}\frac{\bar{z}_{\mathbf{k}\mathbf{G}}^{\nu n}}{\sqrt{\Omega}}Y_{lm}^{*}\left(\widehat{\mathbf{k}+\mathbf{G}}\right)j_{l}\left(\left|\mathbf{k}+\mathbf{G}\right|\left|\mathbf{r}-\mathbf{r}_{\alpha}\right|\right)\exp\left(i\left(\mathbf{k}+\mathbf{G}\right)\cdot\mathbf{r}_{\alpha}\right)\sim\delta_{l,l_{0}}+corrections\label{eq:Proportional}
\end{equation}
\end{widetext}

Where $l_{0}$ is the dominant angular momentum channel of the band
in question. From this we see that only a small number of angular
momentum channels participate much like in lo/LO constructions so
the truncation angular momentum $\bar{l}_{max}$ may be reduced. We
note that $\bar{l}_{max}\rightarrow\bar{l}_{max}^{n\mathbf{k}j\alpha}$
can depend on the iteration $n$, MT sphere $\alpha$ and basis wavefunction
$\mathbf{k}j$.

\subsection{Reducing the number of radial wavefunctions}\label{subsec:Reducing-the-number-1}

In many cases most of the weight of a band in the MT spheres is in
the lowest few angular momentum channels. Then it is permissible to
handle a certain number of angular momentum channels $l_{cut}\leq3,\,\mathrm{or}\,4$
in the EWAPW manner and higher angular momentum channels as in $\left(LA\right)^{2}PW$.
We note that $l_{cut}\rightarrow l_{cut}^{n\mathbf{k}j\alpha}$ can
depend on the iteration $n$, MT sphere $\alpha$ and basis wavefunction
$\mathbf{k}j$. This can increase numerical efficiency.

\subsection{Efficiency of method}\label{subsec:Efficiency-of-method}

We have that the basis size for the EWAPW method is very comparable
to that of APW but the construction of the KS Hamiltonian and charge
density requires a large number of radial integrals see Eqs. (\ref{eq:Overlap_radial})
and (\ref{eq:Definition_Gaunt}) below. This limits the number of
possible windows. However there are many mitigating factors: for a
specific $\mathbf{k}$ point one never needs to consider as many radial
wavefunctions as the total number of bands that are considered relevant
to the windowing construction (see Section \ref{subsec:How-to-generate}
for an example of relevant bands where the occupied and lowest $\mathfrak{M}$
unoccupied bands are relevant so not relevant to the window $\left[E_{N-1,n+1}^{U},E_{N,n+1}^{U}=+\infty\right)$
- in particular one never needs to consider overlap integrals of different
energy levels in the same band even if there are several windows per
band). Furthermore the number of angular momentum channels $\bar{l}_{max}^{n\mathbf{k}j\alpha}$
is smaller then that for LAPW basis and $l_{cut}^{n\mathbf{k}j\alpha}$,
if introduced, can be quite small thereby reducing the number of needed
radial integrals. Furthermore for large scale calculations the diagonalization
time is the dominant contribution to the computer time and should
be very efficient with the EWAPW basis.

\section{Energy Window Full Potential Augmented Plane Waves (EWFAPW)}\label{sec:EWFLAPW}

Here we would like to do full potential calculations for the EWAPW
basis, or EWFAPW. There are many versions of EWAPW, EWLAPW, CEWAPW,
multi radius options, BEE-EWAPW, here we focus on the simplest EWAPW
and do not introduce a $l_{cut}^{n\mathbf{k}j\alpha}$. Extensions
to the more in involved cases are straightforward but tedious. Here,
we break everything (EWFAPW) into simpler pieces - which are easier
to code and reuse many pieces of existing codes - which should be
evident by comparing this calculation with those of refs. \citep{Blaha_1990,Jansen_1984,Hamann_1979,Wei_1985,Wei_1985(2),Wimmer_1981,Mattheis_1986,Blugel_2006}
for FLAPW. Because the wavefunction and its derivative are nearly
continuous near the MT sphere (see Section \ref{subsec:EWAPW-vs.-EWLAPW})
it is reasonably appropriate to use the Hamiltonian from FLAPW (Eq.
(\ref{eq:Main_matrices}) without any boundary terms) appropriate
to LAPW basis set, however the small corrections to the Hamiltonian
due to discontinuity of derivatives of the wavefunction at the MT
spheres have been included for accuracy in the form of a boundary
term. This correction term comes about because the Laplacian term
in the Hamiltonian is not self adjoint when the wavefunctions considered
have derivative discontinuities. This may be corrected with a boundary
term \citep{Loucks_1967,Martin_2020,Marcus_1967,Sjostdet_1999}. Furthermore
the small discontinuity of the wavefunction at the MT spheres (due
to angular momentum truncation) can also be included \citep{Loucks_1967}
in the Hamiltonian. Parts that are not described namely the Weinert
method \citep{Weinert_1981}, total energy as well as how to obtain
$V_{KS}^{n}$ from $\rho^{n}\left(\mathbf{r}\right)$ go over verbatim
in our construction from those corresponding to LAPW basis sets, that
is that part of FLAPW code can be reused directly \citep{Blaha_1990,Jansen_1984,Hamann_1979,Wei_1985,Wei_1985(2),Wimmer_1981,Mattheis_1986,Blugel_2006}.

\subsection{Overlap}\label{subsec:Overlap}

We write that the overlap is a sum of the overlap when restricting
ourselves to the IR plus the sum over the MT spheres:
\begin{equation}
O_{jk}^{\mathbf{k}n}=O_{jk}^{\mathbf{k}nIR}+\sum_{\alpha}O_{jk}^{\mathbf{k}nMT_{\alpha}}\label{eq:Overlap_decomposition}
\end{equation}
Where $O_{jk}^{\mathbf{k}nIR}$ and $O_{jk}^{\mathbf{k}nMT_{\alpha}}$
are the IR and MR contributions respectively. Here 
\begin{align}
O_{jk}^{\mathbf{k}nIR} & =\int_{IR}d\mathbf{r}\Psi_{\mathbf{k}j}^{n+1}{}^{*}\left(\mathbf{r}\right)\Psi_{\mathbf{k}k}^{n+1}\left(\mathbf{r}\right)\nonumber \\
O_{jk}^{\mathbf{k}nMT_{\alpha}} & =\int_{MT_{\alpha}}d\mathbf{r}\Psi_{\mathbf{k}j}^{n+1}{}^{*}\left(\mathbf{r}\right)\Psi_{\mathbf{k}k}^{n+1}\left(\mathbf{r}\right)\label{eq:Overlap_definitions}
\end{align}
We now compute $O_{jk}^{\mathbf{k}nIR}$ and $O_{jk}^{\mathbf{k}nMT_{\alpha}}$
in turn.

\subsubsection{IR Contribution}\label{subsec:IR-Contribution}

We first introduce the function:
\begin{equation}
\Theta^{IR}\left(\mathbf{r}\right)=\left\{ \begin{array}{cc}
1 & \mathbf{r}\in IR\\
0 & \mathbf{r}\in MT
\end{array}\right.\label{eq:Theta_IR}
\end{equation}
Now we Fourier transform the function in Eq. (\ref{eq:Theta_IR}):
\begin{align}
\Theta^{IR}\left(\mathbf{G}\right) & \equiv\frac{1}{\Omega}\int d\mathbf{r}\exp\left(-i\mathbf{G}\cdot\mathbf{r}\right)\Theta^{IR}\left(\mathbf{r}\right)\nonumber \\
 & =\delta_{\mathbf{G},0}-\frac{4\pi}{3\Omega}\sum_{\alpha}\exp\left(-i\mathbf{G}\cdot\mathbf{r}_{\alpha}\right)S_{\alpha}^{3}\frac{j_{1}\left(\left|\mathbf{G}\right|S_{\alpha}\right)}{\left|\mathbf{G}\right|S_{\alpha}}\label{eq:Fourier}
\end{align}
 Next we define an auxiliary overlap function:
\begin{align}
O_{\mathbf{G}\mathbf{G}'}^{\mathbf{k}IR} & =\frac{1}{\Omega}\int_{IR}\exp\left(-i\left(\mathbf{G}-\mathbf{G}'\right)\cdot\mathbf{r}\right)\Theta^{IR}\left(\mathbf{r}\right)\nonumber \\
 & =\Theta^{IR}\left(\mathbf{G}-\mathbf{G}'\right)\label{eq:Overlap_G}
\end{align}
Now we have that the overlap in the IR is given in terms of the auxiliary
overlap function by: 
\begin{equation}
O_{jk}^{\mathbf{k}nIR}=\sum_{\mathbf{G},\mathbf{G}'}\bar{o}_{\mathbf{k}\mathbf{G}}^{nj*}\cdot O_{\mathbf{G}\mathbf{G}'}^{\mathbf{k}IR}\bar{o}_{\mathbf{k}\mathbf{G}'}^{nk}\label{eq:Overlap_IR_final}
\end{equation}

\subsubsection{MT Contribution}\label{subsec:MT-Contribution}

We introduce the matrix of radial integrals
\begin{equation}
O_{E_{i,n}^{V},E_{j,n}^{V}}^{ln\alpha}=\int_{0}^{S_{\alpha}}dr\cdot r^{2}\cdot u_{l}^{n\alpha}\left(r,E_{i,n}^{V}\right)\cdot u_{l}^{n\alpha}\left(r,E_{j,n}^{V}\right)\label{eq:Overlap_radial}
\end{equation}
Now using Eqs. (\ref{eq:Continuous}) and (\ref{eq:Iteration}) we
introduce the constants: 
\begin{equation}
C_{\mathbf{k}lm}^{n\alpha}\left(k\right)=\sum_{\mathbf{G}}\bar{o}_{\mathbf{k}\mathbf{G}}^{nk}\cdot A_{\mathbf{k}\mathbf{G}}^{l\alpha n}\left(\mathcal{E}_{w}^{n}\left(\varepsilon_{\mathbf{k}\nu}^{n-1}\right)\right)\label{eq:Constants}
\end{equation}
with $\nu=k$. Now we have that of the matrix of radial integrals
in Eq. (\ref{eq:Overlap_radial}): 
\begin{equation}
O_{jk}^{\mathbf{k}nIR}=\sum_{lm}C_{\mathbf{k}lm}^{n\alpha*}\left(j\right)O_{\mathcal{E}_{w}^{n}\left(\varepsilon_{\mathbf{k}\nu}^{n-1}\right),\mathcal{E}_{w}^{n}\left(\varepsilon_{\mathbf{k}\mu}^{n-1}\right)}^{ln\alpha}C_{\mathbf{k}lm}^{n\alpha}\left(k\right)\label{eq:Overlap_MT_final}
\end{equation}
Where $j=\mu$ and $k=\nu$.

\subsection{Hamiltonian}\label{subsec:Hamiltonian}

We write that the Hamiltonian is a sum of the overlap when restricting
ourselves to the IR plus the sum over the MT spheres plus a boundary
term. The boundary term \citep{Loucks_1967,Martin_2020} comes about
because we are using wavefunctions which are not continuously differentiable
on the MT spheres whereby the Laplacian becomes non-Hermitian and
a boundary term is needed in order to fix this problem \citep{Loucks_1967,Martin_2020}.
The Hamiltonian is given by: 
\begin{equation}
H_{jk}^{\mathbf{k}n}=H_{jk}^{\mathbf{k}nIR}+\sum_{\alpha}H_{jk}^{\mathbf{k}nMT_{\alpha}}+\sum_{\alpha}H_{jk}^{\mathbf{k}nBd_{\alpha}}\label{eq:Overlap_decomposition-1}
\end{equation}
Where $H_{jk}^{\mathbf{k}nIR}$, $H_{jk}^{\mathbf{k}nMT_{\alpha}}$
and $H_{jk}^{\mathbf{k}nBd_{\alpha}}$ are the IR, MT and boundary
contributions respectively. Here:
\begin{align}
H_{jk}^{\mathbf{k}nIR} & =\int_{IR}d\mathbf{r}\Psi_{\mathbf{k}j}^{n+1}{}^{*}\left(\mathbf{r}\right)\left[-\nabla^{2}+V_{KS}^{n}\left(\mathbf{r}\right)\right]\Psi_{\mathbf{k}k}^{n+1}\left(\mathbf{r}\right)\nonumber \\
H_{jk}^{\mathbf{k}nMT_{\alpha}} & =\int_{MT_{\alpha}}d\mathbf{r}\Psi_{\mathbf{k}j}^{n+1}{}^{*}\left(\mathbf{r}\right)\left[-\nabla^{2}+V_{KS}^{n}\left(\mathbf{r}\right)\right]\Psi_{\mathbf{k}k}^{n+1}\left(\mathbf{r}\right)\label{eq:Hamiltonian_definitions}
\end{align}
 $H_{jk}^{\mathbf{k}nIR}$, $H_{jk}^{\mathbf{k}nMT_{\alpha}}$ and
$H_{jk}^{\mathbf{k}nBd_{\alpha}}$ (given in Eq. (\ref{eq:Boundary})
below) in turn. We now write: 
\begin{equation}
V_{KS}^{n}\left(\mathbf{r}\right)=\left\{ \begin{array}{cc}
\sum_{\mathbf{G}}V_{KS}^{n}\left(\mathbf{G}\right)\exp\left(i\mathbf{G}\cdot\mathbf{r}\right) & \mathbf{r}\in IR\\
\sum_{lm}V_{KS}^{lm\alpha}\left(r\right)Y_{lm}\left(\hat{\mathbf{r}}\right) & \mathbf{r}\in MT_{\alpha}
\end{array}\right.\label{eq:KS_potential}
\end{equation}

\subsubsection{IR Contribution}\label{subsec:IR-Contribution-1}

We now introduce the auxiliary IR Hamiltonian matrix: 
\begin{widetext}
\begin{align}
H_{\mathbf{G}\mathbf{G}'}^{\mathbf{k}nIR} & =\frac{1}{\Omega}\int_{IR}\exp\left(-i\left(\mathbf{k}+\mathbf{G}\right)\cdot\mathbf{r}\right)\left[V_{KS}^{n}\left(\mathbf{r}\right)\Theta^{IR}\left(\mathbf{r}\right)-\nabla^{2}\Theta^{IR}\left(\mathbf{r}\right)\right]\exp\left(i\left(\mathbf{k}+\mathbf{G}'\right)\cdot\mathbf{r}\right)\nonumber \\
 & =\left(V_{KS}^{n}\Theta^{IR}\right)\left(\mathbf{G}\right)+\left(\mathbf{k}+\mathbf{G}'\right)^{2}\Theta^{IR}\left(\mathbf{G}-\mathbf{G}'\right)\label{eq:Overlap_G-1}
\end{align}
\end{widetext}

Where by the Fourier multiplication convolution theorem 
\begin{equation}
\left(V_{KS}^{n}\Theta^{IR}\right)\left(\mathbf{G}\right)=\sum_{\mathbf{G}'}V_{KS}^{n}\left(\mathbf{G}'\right)\Theta^{IR}\left(\mathbf{G}-\mathbf{G}'\right)\label{eq:Convolution}
\end{equation}
Now we have that the main IR Hamiltonian matrix is given by:
\begin{equation}
H_{jk}^{\mathbf{k}nIR}=\sum_{\mathbf{G},\mathbf{G}'}\bar{o}_{\mathbf{k}\mathbf{G}}^{nj*}\cdot H_{\mathbf{G}\mathbf{G}'}^{\mathbf{k}nIR}\bar{o}_{\mathbf{k}\mathbf{G}'}^{nk}\label{eq:H_IR_final}
\end{equation}

\subsubsection{MT Contribution}\label{subsec:MT-Contribution-1}

The MT contribution may be decomposed into a spherical and non spherical
component. We write the spherical harmonics decomposition of the KS
Hamiltonian within a MT sphere (see Eq. (\ref{eq:KS_potential})):
\begin{equation}
-\nabla^{2}+V_{KS}^{n}\left(\mathbf{r}\right)=-\nabla^{2}+\bar{V}_{KS}^{n\alpha}\left(r\right)+\sum_{l\neq0,m}\bar{V}_{KS}^{lmn\alpha}\left(r\right)Y_{lm}\left(\hat{\mathbf{r}}\right)\label{eq:Decomposition}
\end{equation}
Then we write: 
\begin{equation}
H_{jk}^{\mathbf{k}nMT_{\alpha}}=H_{jk}^{\mathbf{k}nSp_{\alpha}}+H_{jk}^{\mathbf{k}nNS_{\alpha}}\label{eq:Decomposition-1}
\end{equation}
Where:
\begin{align*}
H_{jk}^{\mathbf{k}nSp_{\alpha}} & =\int_{MT_{\alpha}}d\mathbf{r}\Psi_{\mathbf{k}j}^{n}{}^{*}\left(\mathbf{r}\right)\left[-\nabla^{2}+\bar{V}_{KS}^{n\alpha}\left(r\right)\right]\Psi_{\mathbf{k}k}^{n}\left(\mathbf{r}\right)\\
H_{jk}^{\mathbf{k}nNS_{\alpha}} & =\int_{MT_{\alpha}}d\mathbf{r}\Psi_{\mathbf{k}j}^{n}{}^{*}\left(\mathbf{r}\right)\sum_{l\neq0,m}\bar{V}_{KS}^{lmn\alpha}\left(r\right)Y_{lm}\left(\hat{\mathbf{r}}\right)\Psi_{\mathbf{k}k}^{n}\left(\mathbf{r}\right)
\end{align*}

\paragraph{Spherical Component}\label{par:Spherical-component}

We first note that: 
\begin{align}
H_{jk}^{\mathbf{k}nSp_{\alpha}} & =\int_{MT_{\alpha}}d\mathbf{r}\Psi_{\mathbf{k}j}^{n}{}^{*}\left(\mathbf{r}\right)\left[-\nabla^{2}+\bar{V}_{KS}^{n\alpha}\left(r\right)\right]\Psi_{\mathbf{k}k}^{n}\left(\mathbf{r}\right)\nonumber \\
 & =\int_{MT_{\alpha}}d\mathbf{r}\Psi_{\mathbf{k}j}^{n}{}^{*}\left(\mathbf{r}\right)E_{i,n}^{V}\Psi_{\mathbf{k}k}^{n}\left(\mathbf{r}\right)\label{eq:Secular-1}
\end{align}
for some $i$. We now introduce the matrix of Hamiltonian radial overlaps:
\begin{equation}
H_{E_{i,n}^{V},E_{j,n}^{V}}^{ln\alpha}\equiv E_{j,n}^{V}\cdot O_{E_{i,n}^{V},E_{j,n}^{V}}^{ln\alpha}\label{eq:Overlap_radial-1}
\end{equation}
We now use Eq. (\ref{eq:Schrodinger}) to write that the spherical
component of the Hamiltonian is given by:
\begin{equation}
H_{jk}^{\mathbf{k}nSp_{\alpha}}=\sum_{lm}C_{\mathbf{k}lm}^{n\alpha*}\left(j\right)H_{\mathcal{E}_{w}^{n}\left(\varepsilon_{\mathbf{k}\nu}^{n-1}\right),\mathcal{E}_{w}^{n}\left(\varepsilon_{\mathbf{k}\mu}^{n-1}\right)}^{ln\alpha}C_{\mathbf{k}lm}^{n\alpha}\left(k\right)\label{eq:Spherical_final}
\end{equation}
Where $j=\nu$ and $k=\mu$.

\paragraph{Non-Spherical Component}\label{par:Non-Spherical-Component}

We now introduce the functions:
\begin{align}
G_{l,l',l"}^{m,m',m"} & =\int Y_{lm}^{*}Y_{l'm'}Y_{l"m"}d\Omega,\nonumber \\
H_{E_{i,n}^{V},E_{j,n}^{V}}^{ll'm',l"n\alpha} & =\int_{0}^{S_{\alpha}}dr\cdot r^{2}\cdot u_{l}^{n\alpha}\left(r,E_{i,n}^{V}\right)\times\nonumber \\
 & \qquad\times\bar{V}_{KS}^{lmn\alpha}\left(r\right)\cdot u_{l"}^{n\alpha}\left(r,E_{j,n}^{V}\right)\label{eq:Definition_Gaunt}
\end{align}
\foreignlanguage{english}{Here $G_{l,l',l"}^{m,m',m"}$ are Gaunt
coefficients \citep{Blugel_2006,Michalicek_2014}. Then the non-spherical
component is given by: }
\selectlanguage{english}%
\begin{widetext}
\begin{equation}
H_{jk}^{\mathbf{k}nNS_{\alpha}}=\sum_{lm,l'\neq0m',l"m"}C_{\mathbf{k}lm}^{n\alpha*}\left(j\right)G_{l,l',l"}^{m,m',m"}H_{\mathcal{E}_{w}^{n}\left(\varepsilon_{\mathbf{k}\nu}^{n-1}\right),\mathcal{E}_{w}^{n}\left(\varepsilon_{\mathbf{k}\mu}^{n-1}\right)}^{ll'm',l"n\alpha}C_{\mathbf{k}l"m"}^{n\alpha}\left(k\right)\label{eq:Final_NS}
\end{equation}
\foreignlanguage{american}{Where $j=\nu$ and $k=\mu$.}
\end{widetext}

\selectlanguage{american}%

\subsubsection{Boundary Contribution}\label{subsec:Boundary-Contribution}

We have that \citep{Loucks_1967,Martin_2020}:
\begin{equation}
H_{jk}^{\mathbf{k}nBd_{\alpha}}=\int_{\partial MT_{\alpha}}d\mathbf{r}\Psi_{\mathbf{k}j}^{n}{}^{*}\left(\mathbf{r}\right)\left[\frac{\partial}{\partial r_{1}}-\frac{\partial}{\partial r_{2}}\right]\Psi_{\mathbf{k}k}^{n}\left(\mathbf{r}\right)\label{eq:Boundary}
\end{equation}
Where $\partial MT_{\alpha}$ is the surface of the $\alpha$'th MT
sphere and $\frac{\partial}{\partial r_{1}}$, $\frac{\partial}{\partial r_{2}}$
refer to derivatives at the inner and outer boundaries. We now introduce
the auxiliary function: 
\begin{widetext}
\begin{align}
H_{\mathbf{G},E_{i,n}^{V},\mathbf{G}'}^{\mathbf{k}lmnBd_{\alpha}} & =\frac{4\pi}{\sqrt{\Omega}}\exp\left(i\left(\mathbf{k}+\mathbf{G}\right)\cdot\mathbf{r}_{\alpha}\right)Y_{lm}^{*}\left(\widehat{\mathbf{k}+\mathbf{G}}\right)i^{l}j_{l}\left(\left|\mathbf{k}+\mathbf{G}\right|S_{\alpha}\right)\times\nonumber \\
 & \times\left[\bar{A}_{\mathbf{k}\mathbf{G}'}^{lm\alpha n}\left(E_{i,n}^{V}\right)\frac{\partial}{\partial r}u_{l\alpha}^{n}\left(E_{i,n}^{V},r\right)_{r=S_{\alpha}}-\frac{4\pi}{\sqrt{\Omega}}\exp\left(i\left(\mathbf{k}+\mathbf{G}'\right)\cdot\mathbf{r}_{\alpha}\right)Y_{lm}^{*}\left(\widehat{\mathbf{k}+\mathbf{G}'}\right)i^{l}\frac{\partial}{\partial r}j_{l}\left(\left|\mathbf{k}+\mathbf{G}'\right|r\right)_{r=S_{\alpha}}\right]\label{eq:Boundary_intermediate}
\end{align}
\end{widetext}

Then we have that:
\begin{equation}
H_{jk}^{\mathbf{k}nBd_{\alpha}}=\sum_{lm}\sum_{\mathbf{G},\mathbf{G}'}\bar{o}_{\mathbf{k}\mathbf{G}}^{nj*}\cdot H_{\mathbf{G},\mathcal{E}_{w}^{n}\left(\varepsilon_{\mathbf{k}\nu}^{n-1}\right),\mathbf{G}'}^{\mathbf{k}lmnBd_{\alpha}}\cdot\bar{o}_{\mathbf{k}\mathbf{G}'}^{nk}\label{eq:Boundary_final}
\end{equation}
Where $j=\nu$.

\subsection{Electron density}\label{subsec:Electron-density}

We would like to write the electron density in the standard form:
\begin{equation}
\rho^{n}\left(\mathbf{r}\right)=\left\{ \begin{array}{cc}
\sum_{\mathbf{G}}\rho^{n}\left(\mathbf{G}\right)\cdot\exp\left(i\mathbf{G}\cdot\mathbf{r}\right) & \mathbf{r}\in IR\\
\sum_{lm}\rho_{lm}^{n\alpha}\left(r\right)Y_{lm}\left(\widehat{\mathbf{r}-\mathbf{r}_{\alpha}}\right) & \mathbf{r}\in MT_{\alpha}
\end{array}\right.\label{eq:Density_decomposition}
\end{equation}
Which is useful for the Weinert method \citep{Weinert_1981} and the
calculation of $V_{KS}^{n}\left(\mathbf{r}\right)$ from $\rho^{n}\left(\mathbf{r}\right)$
\citep{Loucks_1967}.

\subsubsection{IR Component}\label{subsec:IR-Component}

We first introduce the matrices: 
\begin{equation}
Z_{\mathbf{k}\mathbf{G}\nu}^{n}=\sum_{j}\bar{o}_{\mathbf{k}\mathbf{G}}^{nj}\cdot\bar{z}_{\mathbf{k}j}^{\nu n}\label{eq:Planewaves}
\end{equation}
Which give the expansion in terms of plane waves of the eigenfunctions
of the KS Hamiltonian. Then we have that 
\begin{equation}
\rho^{n}\left(\mathbf{G}\right)=2\sum_{\mathbf{k}}\sum_{\nu}f\left(\varepsilon_{\mathbf{k}\nu}^{n}\right)\sum_{\mathbf{G}'}Z_{\mathbf{k}\mathbf{G}\nu}^{n*}Z_{\mathbf{k}\left(\mathbf{G}+\mathbf{G}'\right)\nu}^{n}\label{eq:rho_G}
\end{equation}
The factor of $2$ comes from spin degeneracy.

\subsubsection{MT Component}\label{subsec:MT-Component}

We now introduce the auxiliary matrices: 
\begin{equation}
\mathcal{C}_{\mathbf{k}lm}^{\nu n\alpha}\left(j,r\right)=\bar{z}_{\mathbf{k}j}^{\nu n}\cdot C_{\mathbf{k}lm}^{n\alpha}\left(j\right)u_{l\alpha}^{n}\left(r,\mathcal{E}_{w}^{n}\left(\varepsilon_{\mathbf{k}\mu}^{n-1}\right)\right)\label{eq:Components}
\end{equation}
With $\mu=j$. Then we have that: 
\begin{align}
\rho_{lm}^{n\alpha}\left(r\right) & =2\sum_{\mathbf{k}}\sum_{\nu}f\left(\varepsilon_{\mathbf{k}\nu}^{n}\right)\times\nonumber \\
 & \times\sum_{jk}\sum_{l'm'}\sum_{l"m"}\mathcal{C}_{\mathbf{k}l'm'}^{\nu n\alpha*}\left(j,r\right)\mathcal{C}_{\mathbf{k}l"m"}^{\nu n\alpha}\left(k,r\right)G_{l",l',l}^{m",m',m}\label{eq:Density_spherical}
\end{align}

\section{Discussions \& Conclusions}\label{sec:Conclusions}

In this work we introduced a new method to solve the KS problem once
the form of the correlation and exchange piece of the KS potential
(be it within the Local Density Approximation (LDA), Generalized Gradient
Approximation (GGA), metaGGA etc. \citep{Engel_2011,Dreizler_1990,Tsuneda_2014,Eshcrig_1996,Koch_2001})
has been fixed. In our method, EWAPW, we use the fact that solving
the KS equations is an iterative process and use previous iteration's
eigenstate and eigenvalue information to produce a new basis for each
iteration of the solution of the KS equations (see Section \ref{subsec:Wavefunctions}).
We have argued that no LO/lo or extensions (such as semicore LO, HDLO,
HELO...) would not significantly improve the accuracy of the solution
as we have many linearization energies per occupied band (and a few
for the unoccupied ones). As such we have created a competitor for
LAPW+LO (as well as LAPW+HDLO, LAPW+HELO, LAPW+HDLO+HELO, etc.). However
the basis set size we use is significantly smaller then LO type extensions
as we have one or even less (see Section \ref{subsec:Partial-basis})
eigenstate per reciprocal lattice vector $\mathbf{G}$. This is an
important advantage; indeed typical DFT calculations use 80-100 $\mathbf{G}$
vectors per atom for large MT spheres \citep{Sjostedt_2000,Michalicek_2014},
while LO in the $s$, $p$, $d$, $f$ or even $g$ channels (for
high accuracy) gives an additional 16-25 additional basis functions
per LO band, so if we have LAPW+LO+HDLO+HELO($\times$3) \citep{Michalicek_2013,Michalicek_2014,Kutepov_2021}
(where the ($\times$3) indicates three HELO bands are introduced)
which is likely to have comparable accuracy to EWAPW methods, as it
uses a comparable number of radial wavefunctions per atom, we can
easily need to generate an additional 80-125 basis functions per atom.
Doubling the basis size increase work load for diagonalization by
a factor of \textasciitilde eight so EWAPW is highly advantageous.
Furthermore the Hamiltonian and overlap matrices for EWAPW are nearly
diagonal and sparse giving additional speedup. In other words we have
produced a basis with the efficiency of the APW basis but with accuracy
of LAPW+LO+HDLO+HELO. We would like to note that the efficiency of
a DFT method is mostly determined by the efficiency of the diagonalization
process and the efficiency of setting up the KS Hamiltonian, overlap
matrix and charge distribution. While the EWAPW basis has the efficiency
of diagonalization of APW or better (same basis size, but a sparser
matrix) EWAPW takes comparable time to LAPW+LO+HDLO+HELO to set up
the KS Hamiltonian and overlap matrices as well as the electron density
because of the many radial wavefunctions involved. 

In this paper we focused on a specific method EWAPW, however as we
discussed there are many variations of this method BEE-EWAPW, EWLAPW,
multi radius and CEWAPW options. Furthermore there are many somewhat
different ways of choosing the windowing function; CEWAPW has several
variations and it is possible to introduce a $l_{cut}^{n\mathbf{k}j\alpha}$.
While we have argued that there will not be a very significant change
between all these methods, with respect to the accuracy of the solution
of the KS problem - especially for a large number of windows, in the
future it would still be of great interest to optimize the time requirements
and the accuracy with respect to all these variations of the main
method. Also in the future it would be of interest to combine these
methods (EWAPW and related) with Hartree Fock as in DFT+U \citep{Kotliar_2006},
or Dynamical Mean Field Theory \citep{Georges_1996}, DFT+DMFT \citep{Kotliar_2006},
or Gutzwiller Approximation, DFT+GA \citep{Lanata_2015} to obtain
more accurate results for strongly correlated systems. It would also
be of interest to include magnetism, superconductivity and relativistic
effects for enhanced accuracy, as well as to extend the results to
thin films. Besides expanding the methods it would also be of interest
to expand applications, namely forces (including the Pulay contribution
- which we believe to be very small in this case as the basis is nearly
complete \citep{Yu_1991}), pressure and stress \citep{Klupplenerg_2015,Belbase_2021}.
As well as introduce DFT perturbation theory and study phonons with
various methods \citep{Klupplenerg_2015}. This can be done for a
large variety of options of EWAPW for DFT, DFT+U, DFT+DMFT, DFT+GA.
We would also like to expand EWAPW to Energy Window Muffin Tin Orbitals
(EWMTO \citep{Goldstein_2024(3)}) and to energy window pseudopotential
methods. This would greatly increase the applicability of energy window
methods. Overall it would be of great interest to apply EWAPW to real
materials.

\appendix

\section{ EWLAPW}\label{sec:EWLAPW}

Here we briefly describe EWLAPW, which is similar to EWAPW. We now
introduce:
\begin{widetext}
\begin{equation}
\Psi_{\mathbf{k}\mathbf{G}}^{Ln}\left(E\right)=\left\{ \begin{array}{cc}
\frac{1}{\sqrt{\Omega}}\exp\left(i\left(\mathbf{k}+\mathbf{G}\right)\cdot\mathbf{r}\right) & \mathbf{r}\in IR\\
\left[a_{\mathbf{k}+\mathbf{G}}^{Ll\alpha n}\left(\mathcal{E}_{w}\left(E\right)\right)u_{l\alpha}^{n}\left(r,\mathcal{E}_{w}^{Ln}\left(E\right)\right)+b_{\mathbf{k}+\mathbf{G}}^{Ll\alpha n}\left(\mathcal{E}_{w}\left(E\right)\right)\dot{u}_{l\alpha}^{n}\left(r,\mathcal{E}_{w}^{Ln}\left(E\right)\right)\right]Y_{lm}\left(\widehat{\mathbf{r}-\mathbf{r}_{\alpha}}\right) & \mathbf{r}\in MT_{\alpha}
\end{array}\right.\label{eq:Windowing_auxhilary-1}
\end{equation}
\end{widetext}

Where the wavefunction is continuous and continuously differentiable.
We also note that:
\begin{equation}
\Psi_{\mathbf{k}\mathbf{G}}^{Ln}\left(E\right)=\phi_{\mathbf{k}\mathbf{G}}^{Ln}\left(\mathcal{E}_{w}\left(E\right)\right)\label{eq:Quick}
\end{equation}
Where we have introduced a windowing function $\mathcal{E}_{w}^{Ln}\left(E\right)$
which is given by: 
\begin{equation}
\mathcal{E}_{w}^{Ln}\left(E\right)=\sum_{i=2}^{N}E_{i,n}^{LV}\cdot\left(\Theta\left(E-E_{i-1,n}^{LU}\right)-\Theta\left(E-E_{i,n}^{LU}\right)\right)\label{eq:Window-1}
\end{equation}
Where $E_{1,n}^{LU}=-\infty$, $E_{N,n}^{LU}=+\infty$ and $E_{i-1,n}^{LU}<E_{i,n}^{LV}<E_{i,n}^{LU}$
$\forall i$. Now suppose that for the $n$th step in the solutions
to the KS equations we have a basis of the form: 
\begin{equation}
\Psi_{\mathbf{k}\nu}^{Ln}\left(\mathbf{r}\right)=\left\{ \begin{array}{cc}
\sum_{\mathbf{G}}z_{\mathbf{k}\mathbf{G}}^{L\nu n}\frac{1}{\sqrt{\Omega}}\exp\left(i\left(\mathbf{k}+\mathbf{G}\right)\cdot\mathbf{r}\right) & IR\\
\mathrm{Arbitrary} & MT
\end{array}\right.\label{eq:Basis-1}
\end{equation}
for some $E_{i,n}^{LV}$. Then we that 
\begin{equation}
\Psi_{\mathbf{k}j}^{Ln+1}=\sum_{\mathbf{G}}z_{\mathbf{k}\mathbf{G}}^{L\nu n}\Psi_{\mathbf{k}\mathbf{G}}^{Ln+1}\left(\varepsilon_{\mathbf{k}\nu}^{Ln}\right)\label{eq:Iteration-1}
\end{equation}
for $j=\nu$ and $\varepsilon_{\mathbf{k}\nu}^{Ln}$ are eigenvalues
of the n'th iteration of EWLAPW. The rest is very similar to EWPAW.

\section{Multi Radius options}\label{sec:Multi-Radius-options}

\selectlanguage{english}%
Consider the following wave functions, which are similar to the Soler-Williams
construction \citep{Soler_1989,Soler_1990}: 
\begin{widetext}
\begin{align}
 & \breve{\Psi}_{\mathbf{k}\mathbf{G}}^{n}\left(E\right)=\frac{1}{\sqrt{\Omega}}\exp\left(i\left(\mathbf{k}+\mathbf{G}\right)\cdot\mathbf{r}\right)+\nonumber \\
 & +\sum_{\alpha}\sum_{l,m}Y_{lm}\left(\widehat{\mathbf{r}-\mathbf{r}_{\alpha}}\right)\left[\breve{A}_{\mathbf{k}+\mathbf{G}}^{l\alpha n}\left(E\right)u_{l\alpha}^{n}\left(r,\mathcal{E}_{w}^{n}\left(E\right)\right)-\frac{4\pi}{\sqrt{\Omega}}i^{l}Y_{lm}^{*}\left(\widehat{\mathbf{k}+\mathbf{G}}\right)\exp\left(i\left(\mathbf{k}+\mathbf{G}\right)\cdot\mathbf{r}_{\alpha}\right)j_{l}\left(\left|\mathbf{k}+\mathbf{G}\right|\left|\mathbf{r}-\mathbf{r}_{\alpha}\right|\right)\right]\times\nonumber \\
 & \qquad\qquad\qquad\qquad\qquad\times\Theta\left(S_{\alpha}^{l}\left(\left|\mathbf{k}+\mathbf{G}\right|\right)-\left|\mathbf{r}-\mathbf{r}_{\alpha}\right|\right)\label{eq:SAPW}
\end{align}
\end{widetext}

Here we have subtracted out the part of the plane wave inside the
MT sphere, in the second term of the second line of Eq. (\ref{eq:SAPW}),
that is decomposed it as in Eq. (\ref{eq:Famous}) into Bessel functions
times spherical harmonics and subtracted them piece by piece from
the solutions of the exact radially symmetric KS equations. We note
that these wavefunctions are well defined even when the MT spheres,
with radii $S_{\alpha}^{l}\left(\left|\mathbf{k}+\mathbf{G}\right|\right)$,
are such that the MT spheres overlap. Notice that there are different
radii for different angular momentum channels and different magnitudes
of wavevector $\left|\mathbf{k}+\mathbf{G}\right|$. We also choose
the coefficients $\breve{A}_{\mathbf{k}+\mathbf{G}}^{l\alpha n}$
to be given by a relation similar to Eq. (\ref{eq:Continuous}) such
that the wave function is continuous at $S_{\alpha}^{l}\left(\left|\mathbf{k}+\mathbf{K}\right|\right)$.
Now we choose $S_{\alpha}^{l}\left(\left|\mathbf{k}+\mathbf{K}\right|\right)$
such that the wave function is also continuously differentiable at
$S_{\alpha}^{l}\left(\left|\mathbf{k}+\mathbf{K}\right|\right)$ (this
may be done by a root finding procedure \citep{Goldstein_20024}).
We now repeat the arguments in Section \ref{sec:Main-Explanations}
with $\Psi_{\mathbf{k}\mathbf{G}}^{n}\left(E\right)\rightarrow\breve{\Psi}_{\mathbf{k}\mathbf{G}}^{n}\left(E\right)$
to obtain a basis set procedure. We note that the linearization error
goes as $\sim\left(\varepsilon_{\mathbf{k}\nu}^{n+1}-\mathcal{E}_{w}\left(\varepsilon_{\mathbf{k}\nu}^{n}\right)\right)^{4}$
\citep{Singh_2006,Martin_2020,Goldstein_20024} instead of $\sim\left(\varepsilon_{\mathbf{k}\nu}^{n+1}-\mathcal{E}_{w}\left(\varepsilon_{\mathbf{k}\nu}^{n}\right)\right)^{2}$
despite having APW basis set that is no terms proportional to $\dot{u}_{l\alpha}^{n}\left(r,\mathcal{E}_{w}^{n}\left(E\right)\right)$
\citep{Goldstein_20024}. Indeed by matching the derivative we have
essentially studied the case when there is $\dot{u}_{l\alpha}^{n}\left(r,\mathcal{E}_{w}^{n}\left(E\right)\right)$
(it is included) and we found that the coefficient in front of it
is zero at that radius \citep{Goldstein_20024}.

\section{CEWAPW}\label{sec:CEWPAW}

\selectlanguage{american}%
CEWAPW is based on the following wavefunctions: 
\begin{widetext}
\begin{align}
\tilde{\Phi}_{\mathbf{k}\mathbf{G}}^{n}\left(E\right) & =\sum_{i=3}^{N-1}\left[\frac{\phi_{\mathbf{k}\mathbf{G}}^{n}\left(E_{i-1,n}^{CU}\right)\left(E_{i,n}^{CU}-E\right)+\phi_{\mathbf{k}\mathbf{G}}^{n}\left(E_{i,n}^{CU}\right)\left(E-E_{i-1,n}^{CU}\right)}{E_{i,n}^{CU}-E_{i-1,n}^{CU}}\right]\cdot\left(\Theta\left(E-E_{i-1,n}^{CU}\right)-\Theta\left(E-E_{i,n}^{CU}\right)\right)\nonumber \\
 & +\phi_{\mathbf{k}\mathbf{G}}^{n}\left(E_{2,n}^{CU}\right)\Theta\left(E_{2,n}^{CU}-E\right)+\phi_{\mathbf{k}\mathbf{G}}^{n}\left(E_{N-1,n}^{CU}\right)\Theta\left(E-E_{N-1,n}^{CU}\right)\label{eq:Window_interpolate}
\end{align}
\end{widetext}

Where $E_{1,n}^{CU}=-\infty$, $E_{N,n}^{CU}=+\infty$ and $E_{i-1,n}^{CU}<E_{i,n}^{CU}$
$\forall i$. These wavefunctions continuously interpolate between
the various wavefunctions $\phi_{\mathbf{k}\mathbf{G}}^{n}\left(E_{i,n}^{CU}\right)$.
One then replaces $\tilde{\Phi}_{\mathbf{k}\mathbf{G}}^{n}\left(E\right)\leftrightarrow\Phi_{\mathbf{k}\mathbf{G}}^{n}\left(E\right)$
in Section \ref{subsec:Wavefunctions} and proceeds to build a basis
of near eigenstates. Alternatively one could use even more sophisticated
interpolation schemes and write $\hat{\Phi}_{\mathbf{k}\mathbf{G}}^{n}\left(E\right)=$:
\begin{align}
 & \sum_{i=2}^{N-1}\phi_{\mathbf{k}\mathbf{G}}^{n}\left(E_{i,n}^{CU}\right)\prod_{j\neq i}\frac{\left(E-E_{j,n}^{CU}\right)}{\left(E_{i,n}^{CU}-E_{j,n}^{CU}\right)}\Theta\left(E_{N-1,n}^{CU}-E\right)+\nonumber \\
 & +\phi_{\mathbf{k}\mathbf{G}}^{n}\left(E_{N-1,n}^{CU}\right)\Theta\left(E-E_{N-1,n}^{CU}\right)\label{eq:Interpolation}
\end{align}
One then replaces $\hat{\Phi}_{\mathbf{k}\mathbf{G}}^{n}\left(E\right)\leftrightarrow\Phi_{\mathbf{k}\mathbf{G}}^{n}\left(E\right)$
in Section \ref{subsec:Wavefunctions} and proceeds to build a basis
of near eigenstates. In general any reasonable linear interpolation
method, that sends the constant sequence to a constant function function,
would do. We note that: 
\begin{align}
 & \sum_{i=2}^{N-1}\frac{1}{\sqrt{\Omega}}\exp\left(i\left(\mathbf{k}+\mathbf{G}\right)\cdot\mathbf{r}\right)\prod_{j\neq i}\frac{\left(E-E_{j,n}^{CU}\right)}{\left(E_{i,n}^{CU}-E_{j,n}^{CU}\right)}\nonumber \\
 & =\frac{1}{\sqrt{\Omega}}\exp\left(i\left(\mathbf{k}+\mathbf{G}\right)\cdot\mathbf{r}\right)\label{eq:IR}
\end{align}
So that 
\begin{equation}
\hat{\Phi}_{\mathbf{k}\mathbf{G}}^{n}\left(E\right)=\frac{1}{\sqrt{\Omega}}\exp\left(i\left(\mathbf{k}+\mathbf{G}\right)\cdot\mathbf{r}\right)\label{eq:Good}
\end{equation}
for $\mathbf{r}\in IR$.
\selectlanguage{english}%

\section{Adjusting the basis set size with iteration (review)}\label{sec:Adjusting-the-basis}

\selectlanguage{american}%
Most current DFT calculations for crystalline solids introduce a discrete
basis set with the basis set often being indexed by $\mathbf{k},\mathbf{G}$
and for practical reasons is limited to $\mathbf{G}<\mathbf{G}_{max}$,
while $\mathbf{k}$ is on a discrete grid. Typically several hundred
$\mathbf{k}$ points are taken and roughly 80-100 reciprocal lattice
vectors, $\mathbf{G}$ points, per atom \citep{Michalicek_2014,Singh_2006}
are used for the case of large MT spheres. In many DFT calculations
the dominant contribution to computational time is solving Eq. (\ref{eq:Secular})
for a total of $\mathcal{N}$ times till convergence with each run
taking $\sim\mathbf{G}_{max}^{9}$ time for a total of $\sim\mathcal{N}\mathbf{G}_{max}^{9}$
computations. Here $\mathcal{N\sim}20$ for many situations \citep{Michalicek_2014}.
Now at each step from numerical simulations it is known that the error
from an exact solution the KS equations is given by \citep{Singh_2006,Michalicek_2014,Woods_2019,Sjostedt_2000}:
\begin{equation}
\Delta\varepsilon\sim\left[\varepsilon_{n}\exp\left(-An^{\alpha}\right)+\varepsilon_{\mathbf{G}}\exp\left(-B\mathbf{G}_{max}^{\beta}\right)\right]\label{eq:Extremization_error}
\end{equation}
For $\varepsilon_{n}\sim\varepsilon_{\mathbf{G}}\lesssim$1 Ry. Furthermore
in many cases $\alpha\cong\beta\cong1$ - \citep{Michalicek_2014,Woods_2019}.
This shows that the scaling of the computation time $\sim\mathcal{N}\mathbf{G}_{max}^{9}$
is very wasteful as for small $n$ there is no point in using as large
a $\mathbf{G}_{max}$. Here we propose to introduce a $\mathbf{G}_{max}\left(n\right)$
where 
\begin{equation}
B\mathbf{G}_{max}^{\beta}\left(n\right)\sim An^{\alpha}\Rightarrow\mathbf{G}_{max}\left(n\right)\sim\left(\frac{A}{B}\right)^{1/\beta}n^{\alpha/\beta}\label{eq:Running_cutoff}
\end{equation}
We also want $\mathbf{G}_{max}\left(\mathcal{N}\right)=\mathbf{G}_{max}$.
This gives us a first approximation that 
\begin{equation}
\mathbf{G}_{max}\left(n\right)\sim\mathbf{G}_{max}\left(\frac{n}{\mathcal{N}}\right)^{\alpha/\beta}\label{eq:Scaling}
\end{equation}
As such $\mathbf{G}_{max}\left(\mathcal{N}-n\right)=\mathbf{G}_{max}\left(\frac{\mathcal{N-}n}{\mathcal{N}}\right)^{\alpha/\beta}=\mathbf{G}_{max}\left(1-\frac{n}{\mathcal{N}}\right)^{\alpha/\beta}=\mathbf{G}_{max}\exp\left(-\frac{\alpha}{\beta}\frac{n}{\mathcal{N}}\right)$
computational time will then scale as: 
\begin{align}
T & \sim\sum_{n}\mathbf{G}_{max}^{9}\left(\mathcal{N}-n\right)\sim\int_{-\infty}^{0}dt\mathbf{G}_{max}^{9}\exp\left(-9\frac{\alpha}{\beta}\frac{t}{\mathcal{N}}\right)\nonumber \\
\Rightarrow & T\sim\mathbf{G}_{max}^{9}\left[\frac{\mathcal{N\beta}}{9\alpha}+1\right]\label{eq:Time}
\end{align}
Where we have estimated the last iteration loop more carefully as
it is often important (hence $\frac{\mathcal{N\beta}}{9\alpha}\rightarrow\frac{\mathcal{N\beta}}{9\alpha}+1$).
This leads to a computational improvement of the scale of $\sim\frac{\mathcal{N}}{\frac{\mathcal{N\beta}}{9\alpha}+1}$
on top of those described in the main text. We note that it is also
possible to modify the $\mathbf{k}$ space mesh to be more refined
for large $n$ thereby saving further computer time, this works with
a variety of solid state DFT methods including EWAPW. Furthermore,
for EWAPW it is possible to increase the number of windows with convergence
iterations thereby saving on EWFAPW time on top of the previous time
saving modifications. We further note that for calculations with LO/lo
type wavefunctions (likely not relevant to EWAPW) it can be advantageous
to increase the amount of LO/lo basis elements with convergence iterations.
\selectlanguage{english}%

\end{document}